\documentclass[10pt,final,journal,letterpaper,twoside,twocolumn]{IEEEtran}
\usepackage{amsfonts}
\usepackage{amssymb}
\usepackage{amsmath}
\usepackage{psfrag}
\usepackage[dvips]{graphicx}
\usepackage{cite}
\usepackage{balance}
\usepackage{url}

\DeclareMathOperator{\erfc}{erfc}

\newcommand\abs[1]{\lvert#1\rvert}
\newcommand\re[1]{\mathrm{Re}\{#1\}}
\newcommand\im[1]{\mathrm{Im}\{#1\}}
\newcommand\av[1]{\mathrm{E}\{#1\}}

\begin{document}

\markboth{Published in IEEE Transactions on Wireless Communications}{Boulogeorgos \MakeLowercase{\textit{et al.}}: I/Q-Imbalance Self-Interference Coordination}

\title{I/Q-Imbalance Self-Interference Coordination}

\author{Alexandros--Apostolos~A.~Boulogeorgos,~\IEEEmembership{Student~Member,~IEEE},
Vasileios~M.~Kapinas,~\IEEEmembership{Member,~IEEE},
Robert~Schober,~\IEEEmembership{Fellow,~IEEE}, and George~K.~Karagiannidis,~\IEEEmembership{Fellow,~IEEE}

\thanks{Copyright (c) 2015 IEEE. Personal use of this material is permitted. However, permission to use  this material for any other purposes must be obtained from the IEEE by  sending a request to pubs-permissions@ieee.org.}
\thanks{Manuscript received July 15, 2015; revised December 18, 2015; accepted February 20, 2016. Date of publication xxxxx, xx xxxx; date of current version xxxxx, xx xxxx. The associate editor coordinating the review of this paper and approving it for publication was Prof. Mohamed-Slim Alouini.}%.
\thanks{A.--A.~A.~Boulogeorgos, V.~M. Kapinas, and G.~K. Karagiannidis are with the Department of Electrical and Computer Engineering, Aristotle University of Thessaloniki, 54636, Thessaloniki, Greece (e-mail: \{ampoulog, kapinas, geokarag\}@auth.gr).}%
\thanks{R. Schober is with the Institute for Digital Communications, Universit\"at Erlangen-N\"{u}rnberg, Erlangen 91058, Germany (e-mail: schober@lnt.de).}
\thanks{Digital Object Identifier 10.1109/TWC.2016.2535441}
}

\maketitle

\begin{abstract}
In this paper, we present a novel low-complexity scheme, which improves the performance of single-antenna multi-carrier communication systems, suffering from in-phase and quadrature (I/Q)-imbalance (IQI) at the receiver. We refer to the proposed scheme as \textit{I/Q-imbalance self-interference coordination (IQSC)}.
IQSC does not only mitigate the detrimental effects of IQI, but, through appropriate signal processing, also coordinates the self-interference terms produced by IQI in order to achieve second-order frequency diversity. However, these benefits come at the expense of a reduction in transmission rate. More specifically, IQSC is a simple transmit diversity scheme that improves the signal quality at the receiver by elementary signal processing operations across symmetric (mirror) pairs of subcarriers. Thereby, the proposed transmission protocol has a similar complexity as Alamouti's space-time block coding scheme and does not require extra transmit power nor any feedback. To evaluate the performance of IQSC, we derive closed-form expressions for the resulting outage probability and symbol error rate. Interestingly, IQSC outperforms not only existing IQI compensation schemes but also the ideal system without IQI for the same spectral efficiency and practical target error rates, while it achieves almost the same performance as ideal (i.e., IQI-free) equal-rate repetition coding. Our findings reveal that IQSC is a promising low-complexity technique for significantly increasing the reliability of low-cost devices that suffer from high levels of~IQI.
\end{abstract}

\begin{IEEEkeywords}
Direct-conversion architecture, hardware imperfection, I/Q imbalance, I/Q imbalance compensation, image rejection ratio, mirror-frequency diversity, multi-carrier communication systems, radio frequency (RF) impairments, self-interference coordination, transmit diversity.
\end{IEEEkeywords}

\section{Introduction}\label{S:Intro}

The fast evolution of wireless communication systems is driving the design and implementation of modern flexible and software-configurable radio transceivers \cite{B:CMOS,A:Blind_Fr_Dependent_IQI_Comp_for_DC_RX}. By definition, flexible radios are characterized by the ability to operate over multiple-frequency bands, the support of different types of waveforms, and the compatibility with current and future air interface technologies. The terms multi-mode and multi-band are commonly used in this context. Furthermore, next-generation wireless networks are expected to support high data-rate applications and services that require efficient and low-cost wideband radio designs for the mobile terminal~\cite{A:Circularity_based_IQI_Comp_in_wideband_DC_RXs}.

In this context, the well-known direct-conversion architecture (DCA) has become instrumental for realizing compact, low-power, and low-cost transceiver designs for wideband radio \cite{A:SDR_4G}. In direct-conversion receivers (DCRs), quadrature mixing is used, which theoretically provides infinite attenuation of the image band and removes the need for analog image-rejection filtering. However, in practice, the DCA is sensitive to imperfections of the analog radio frequency (RF) front-end sections of the transceiver due to fundamental physical limitations \cite{B:Ravazi,A:Dig_Comp_IQI_effects_in_STCT,DBLP:journals/corr/BoulogeorgosCK15}. An indicative example of such limitations is the so-called in-phase and quadrature (I/Q) imbalance (IQI), which stems from the unavoidable amplitude and phase differences between the physical analog in-phase ($I$) and quadrature ($Q$) signal paths. This problem arises mainly because of the finite tolerances of the capacitors and resistors used in the implementation of the analog RF front-end components. Although a perfectly balanced quadrature down-conversion corresponds to a pure frequency translation, IQI introduces a frequency translation that results in a mixture of image and desired signals. In more detail, I/Q mismatches decrease the theoretically infinite image rejection ratio (IRR) of the receiver down to $20-40$ dB, resulting in crosstalk or interference between mirror frequencies \cite{A:Circularity_based_IQI_Comp_in_wideband_DC_RXs,DAC:DAC561}. Consequently, IQI degrades the effective signal-to-interference-plus-noise ratio (SINR) and causes performance degradation. The impact of IQI is more severe in systems employing high-order modulations and high coding rates, such as Wireless Local Area Networks (WLANs), Worldwide Interoperability for Microwave Access (WiMAX), Long-Term Evolution (LTE), and Digital Video Broadcasting (DVB), among other standards \cite{IMT_advanced_relay_standarts}. Hence, effective IQI compensation is essential for the design of high data-rate communication systems employing the~DCA.

\subsection{Related Work}\label{SS:Intro_RW}

The effects of RF imperfections in general were studied in several works~\cite{B:Schenk-book,
B:wenk2010mimo,
RF_impairments_generalized_model,
A:A_new_look_at_dual_hop_relaying_performance_limits_with_hw_impairments,
Performance_of_MIMO_OFDM_Systems_with_Phase_Noise_at_Transmit_and_Receive_Antennas,
Sensitivity_of_Spectrum_Sensing_to_RF_impairments,
Cyclostationary_Sensing_of_OFDM_RF_impairments,
ED_PHN,
Boul1506:Energy},
while performance degradation due to IQI in particular was investigated in~\cite{A:Analysis_and_comp_of_IQI_in_MIMO_TX_RX_diversity_systems,
A:IQI_SINR_ODFM,
Boul1508:Effects,
C:IQI_BER,
IQI_IRR_practical_values,
C:Compensation_HPA_IQI_MIMO_beamforming,
Performance_impact_of_IQ_mismatch_in_direct_conversion_MIMO_OFDM_transceivers,
A:Impact_of_Transceiver_IQI_on_Transmit_Diversity_of_Beamforming_OFDM_Systems,
A:IQI_TX_RX_AF_Alouini,
A:IQI_in_AF_Nakagami_m,
A:OFDM_OR_IQI,
A:IQI_in_Two_Way_AF_relaying,
A:Impairments_on_AF_relaying,
A:Energy_Detection_under_IQI}.
For instance, the authors in \cite{A:IQI_SINR_ODFM} derived an exact expression for the SINR in orthogonal frequency division multiplexing (OFDM) systems impaired by IQI, assuming that the channel of each subcarrier and its image are uncorrelated. In \cite{C:IQI_BER}, the performance of OFDM systems employing $M$-ary quadrature amplitude modulation (QAM) was studied in the presence of IQI in terms of the error vector magnitude, which is a modulation quality measure used to evaluate the effects of imperfections in digital communication systems.
In \cite{IQI_IRR_practical_values}, the impact of IQI caused by a low pass filter mismatch was illustrated and the importance of IQI compensation was highlighted. Furthermore, the authors of \cite{A:IQI_TX_RX_AF_Alouini,A:IQI_in_AF_Nakagami_m,A:OFDM_OR_IQI,A:IQI_in_Two_Way_AF_relaying} analyzed the performance of relaying systems in the presence of IQI. The impact of IQI in cognitive radio systems was analyzed in \cite{Sensitivity_of_Spectrum_Sensing_to_RF_impairments,A:Energy_Detection_under_IQI, Boul1506:Energy}, where it was shown that, in a multi-channel environment, IQI increases the false-alarm probability significantly and considerably limits the spectrum sensing capabilities of energy detectors compared to the ideal RF receiver case.

Various approaches have been proposed so far to eliminate, compensate, and mitigate the effects of IQI using baseband signal processing techniques, see \cite{A:Blind_Fr_Dependent_IQI_Comp_for_DC_RX,
A:Circularity_based_IQI_Comp_in_wideband_DC_RXs,
A:On_the_connection_of_IQI_and_Channel_eq_in_CD_TXRX,
A:Joint_Estimation,
A:Dig_Comp_IQI_effects_in_STCT,
A:Advanced_methods_IQI_comp,
A:Signal_path_opt_in_SDR,
A:Frequency_offset_and_IQI_comp_for_DC_RX,
Pilot_Design_for_Channel_Estimation_of_MIMO_OFDM_Systems_with_Frequency_Dependent_IQI,
A:Digital_Compensation_of_Frequency_Dependent_IQI_in_OFDM_Systems_Under_High_Mobility,
A:IQI_compensation_broadband_direct_conversion_TX_Valkama,
C:Compensation_HPA_IQI_MIMO_beamforming,
A:Analysis_and_Compensation_of_IQI_in_AF_Cooperative_Systems,
A:Comp_schemes_and_anal_IQI_OFDM_RX}, and references therein.
For example, in \cite{Pilot_Design_for_Channel_Estimation_of_MIMO_OFDM_Systems_with_Frequency_Dependent_IQI}, the authors proposed a number of pilot designs for channel estimation in OFDM systems in the presence of I/Q mismatches at both the transmitter (TX) and the receiver (RX). Moreover, estimation-based system-level algorithms, including least square equalization, adaptive equalization, and post-fast Fourier transform least square, were proposed in \cite{A:Comp_schemes_and_anal_IQI_OFDM_RX} to compensate the distortions caused by IQI. Furthermore, blind (non-data-aided) digital signal processing-based compensation of IQI for wideband multi-carrier systems was studied in \cite{A:Blind_Fr_Dependent_IQI_Comp_for_DC_RX,A:Circularity_based_IQI_Comp_in_wideband_DC_RXs,
A:Dig_Comp_IQI_effects_in_STCT,A:Advanced_methods_IQI_comp,
Boul1506:Energy}. Specifically, in \cite{A:Dig_Comp_IQI_effects_in_STCT} a digital compensation method was proposed for multiple-input multiple-output (MIMO) systems employing space-time block coding, which is based on the algebraic properties of the received signal combined with a suitable pilot structure, while interference cancellation-based and blind source separation-based compensation methods were presented in~\cite{A:Advanced_methods_IQI_comp}.

All previously mentioned works deal with IQI as a source of impairment that should be compensated. In contrast to this approach, IQI at the TX may also be seen as a source of diversity, due to the TX-induced mirror-frequency interference. This diversity can be fully exploited via joint maximum likelihood (ML) detection of the signals received in the mirror subcarriers, or partially exploited by other sub-optimal nonlinear detection techniques such as successive interference cancellation (SIC), as was demonstrated experimentally for OFDM in \cite{C:Div_Gain_TX_IQI}, and later confirmed in \cite{C:Diversity_Gain_IQI_analysis}. Still, when weighed against the implementation complexity of nonlinear RXs, the small achievable signal-to-noise ratio ($\rm{SNR}$) improvement may prove to be too expensive \cite{PhD:Antilla}.
Moreover, as pointed out in several prior works including \cite{C:Div_Gain_TX_IQI}, RX IQI is detrimental for the outage and error performance of wireless communication systems, regardless of the detector that is used. The reason for this is that RX IQI affects both the received signal and the noise; hence, it is commonly believed that RX IQI should be compensated \cite{A:Blind_Fr_Dependent_IQI_Comp_for_DC_RX,A:Circularity_based_IQI_Comp_in_wideband_DC_RXs,
A:Dig_Comp_IQI_effects_in_STCT,A:Advanced_methods_IQI_comp,A:Analysis_and_comp_of_IQI_in_MIMO_TX_RX_diversity_systems,
A:Analysis_and_Compensation_of_IQI_in_AF_Cooperative_Systems,A:OFDM_OR_IQI}. However, to the best of the authors' knowledge, no solution has been proposed so far that achieves a diversity gain in the presence of RX IQI.

\subsection{Motivation and Contribution}\label{SS:Intro_Cont}

From an implementation point of view, DCA is a promising approach to realize low-cost highly integrated wireless equipment. Although DCRs avoid the main drawbacks of other RX architectures, the insufficient image rejection due to IQI is a major concern. For instance, in the case of using a non-zero intermediate frequency (IF), the image signal can be up to $50-100$ dB stronger than the desired one \cite{A:Advanced_methods_IQI_comp}. Thus, in such situations, the $20-40$ dB attenuation provided by the quadrature down-conversion alone is clearly insufficient. Furthermore, with wideband modulated communication waveforms and high-order symbol alphabets, IQI has a tremendous impact on the demodulated signal quality and can severely degrade the RX performance, if not taken properly into account.
Notice also that, although the distortion caused by IQI resembles to some extent ordinary inter-symbol interference (ISI), it cannot be properly mitigated using ordinary equalization techniques due to its special structure~\cite{A:Dig_Comp_IQI_effects_in_STCT}.

In this paper, we propose a novel low-complexity technique, which we refer to as \textit{I/Q-imbalance} \textit{self}-\textit{interference} \textit{coordination} \textit{(IQSC)}, which significantly increases the performance of single-antenna multi-carrier communication systems suffering from IQI at the RX, by coordinating the self-interference caused by IQI. In contrast to the IQI compensation approach, IQSC does not only eliminate the effects of IQI, but, through signal processing, also coordinates the self-interference terms produced by IQI to achieve frequency diversity, which we refer to as \textit{mirror}-\textit{frequency} \textit{diversity} \textit{(MFD)}.
In other words, IQSC is a low-complexity transmit diversity scheme, which improves the signal quality at the RX by simple signal processing operations across symmetric subcarrier pairs. IQSC achieves a diversity order of two, i.e., the same diversity order as maximal-ratio combining (MRC) with two RX antennas or Alamouti's space-time block code with two TX antennas \cite{A:Alamouti}. Notably, by applying IQSC in DCA systems, the IQI is not only compensated but is no longer harmful for the system's performance. Furthermore, the proposed transmission protocol requires neither extra transmit power nor any feedback from the RX to the TX, while its computational complexity is similar to that of Alamouti's space-time block coding scheme. However, the exploitation of the extra degrees of freedom (DoF) offered by IQI to achieve two-fold transmit diversity comes at the expense of a reduction in transmission rate.
In particular, the encoding process at the TX requires two consecutive time intervals (or time-slots) for transmission of each data block. Finally, an alternative IQSC \textit{(A-IQSC)} technique is also presented, which achieves similar outage and error performance as IQSC, with the same computational complexity and~rate.

To confirm the effectiveness of the proposed method, we derive closed-form expressions for the outage probability and the symbol error rate (SER) of IQSC, and compare its performance (with respect to these two metrics) with two baseline systems; namely an ideal system without IQI, called \textit{ideal RF front-end}, and a system with uncompensated IQI in the RF front-end, called \textit{IQI RF front-end}, which correspond to the best case and the conventional transmission scenarios, respectively. Our results demonstrate the superior reliability (which comes at the expense of sacrificing throughput) of IQSC compared to the baseline systems. Surprisingly, in medium-to-high $\rm{SNRs}$, the proposed scheme outperforms both baseline schemes in terms of error performance even for the same spectral efficiency. As another means to assess its error performance fairly, we also compare IQSC with repetition coding (RC) and the frequency-time block code (FTBC) proposed in \cite{A:FTBC}, which both have the same rate as IQSC. Again, we observe that IQSC outperforms both RC and FTBC with uncompensated IQI, referred to as \textit{IQI RF front-end with RC} and \textit{FTBC}, respectively, while it has a similar performance as the \textit{ideal RF front-end with RC} and \textit{FTBC}, i.e., RC and FTBC without IQI in the RF front-end.

\subsection{Organization and Notations}\label{SS:Intro_Not}

%\subsubsection*{Organization}
The remainder of this paper is organized as follows. The equivalent complex baseband signal representation of a multi-carrier communication system with IQI at the RX is presented in Section~\ref{S:SM}. In Section~\ref{S:IQSC}, the proposed IQSC encoding at the TX and the associated combining at the RX, as well as an alternative IQI coordination scheme, namely A-IQSC, are presented in detail. A performance analysis of IQSC in terms of outage probability and SER is provided in Section~\ref{S:Perf}. A point-to-point comparison of IQSC with a system employing RC, under the same bandwidth and power constraints, is given at the end of the same section. In Section~\ref{S:Sim}, we verify our theoretical analysis by computer simulations, confirming that IQSC is a robust technique for multi-carrier transmission under IQI. In Section~\ref{S:Talk}, the main merits and drawbacks of IQSC are outlined, followed by some discussion regarding the new concept of MFD. Finally, Section~\ref{S:End} concludes the paper by summarizing our main~findings.

%\subsubsection*{Notations}
In this paper, $z^\ast$ and $\abs{z}$ denote the complex conjugate and the absolute value of complex number $z$, respectively, while $\re{z}$ and $\im{z}$ represent its real and imaginary part, respectively. Additionally, $\{a_k\}_{k=1}^{n}$ is a shorthand notation for set $\{a_1,\dotsc,a_n\}$, while $\left(a_1,\dotsc,a_n\right)$ defines a sequence of $n$ terms. $\av{z}$ gives the expected value of random variable (RV) $z$. Finally, $\erfc\left(x\right)=\frac{2}{\sqrt{\pi}}\int_{x}^{\infty}\exp\left(-t^{2}\right)dt$ stands for the complementary error function~\cite{Integrals_and_series}.

\section{System and Signal Model}\label{S:SM}

We start this section by considering the baseband signal model of a system without IQI, which will be referred to as the ideal RF front-end. Having this as a reference, we present the practical IQI signal model of multi-carrier DCRs, assuming a single antenna at both the TX and the RX, perfect channel estimation at the RX, and no channel state information (CSI) at the~TX.

\subsection{Ideal RF Front-End}\label{SS:SM_Ideal}

We consider a multi-carrier system with $2K$ RF subcarriers and assume down-conversion to baseband using the wideband direct-conversion principle. For notational convenience, we denote the set of these subcarriers~as
\begin{align}
\{-K,\cdots,-1,1\cdots,K\}=\{k\}_{k=-K}^{K},
\end{align}
and the set of data symbols loaded to them~as
\begin{align}
S_K = \left\{s(-K),\dotsc,s(-1),s(1),\dotsc,s(K)\right\}=\{s(k)\}_{k=-K}^{K}.\label{Eq:symbols_K}
\end{align}
We further assume flat fading on each subcarrier, and that the RF front-ends of both the TX and the RX are perfect, i.e., no IQI is present in the system.

The received signal is passed through various front-end stages, including filtering, amplification, analog I/Q demodulation (down-conversion), and sampling. The baseband equivalent received signal in subcarrier $k$~is
\begin{align}
r_\text{id}(k) = h(k) s(k) + n(k),\label{Eq:r_ideal}
\end{align}
where fading gain $h(k)$ is modeled as a zero-mean complex Gaussian process of unit variance, and $n(k)$ represents circularly symmetric additive white Gaussian noise (AWGN) with power spectral density~$N_0$.

\subsection{I/Q Imbalance Model}\label{SS:SM_IQI}

\begin{figure}
\psfrag{LPF}[][][1]{LPF}%
\psfrag{A/D}[][][1]{A/D}%
\psfrag{j}[][][1]{$j$}%
\psfrag{x}[][][1]{$\cos(\omega_\text{LO}t)$}%
\psfrag{y}[][][1]{$-\epsilon\sin(\omega_\text{LO}t+\phi)$}%
\psfrag{r(t)}[][][1]{$r(t)$}%
\centering\includegraphics[width=1.0\linewidth,trim=0 0 0 0,clip=false]{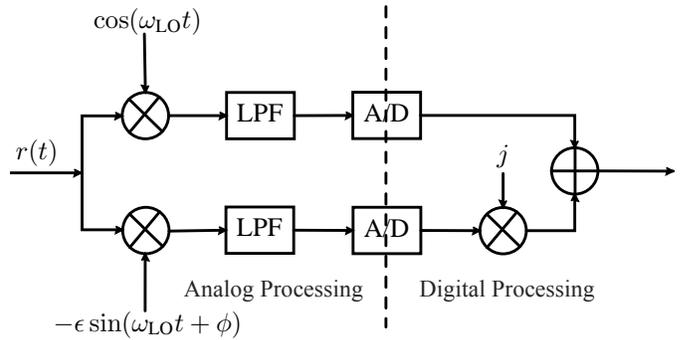}
\caption{IQI model. In the block diagram, LPF and A/D denote the low pass filter and the analog-to-digital converter, respectively.}
\label{fig:IQI_model}
\end{figure}

In general, all analog components of the RX's $I$ and $Q$ branches contribute to the IQI effect. One obvious source of imbalance is the I/Q mixer stage, which is typically modeled in terms of the I/Q local oscillator (LO) signals, namely $\cos(\omega_\text{LO}t)$ and $-\epsilon\sin(\omega_\text{LO}t+\phi)$, where $\epsilon$ and $\phi$ represent the relative amplitude and phase imbalances, respectively, and $\omega_\text{LO}=2\pi f_\text{LO}$ is the center frequency, as shown in Fig.~\ref{fig:IQI_model}. As a consequence, the total imbalance between the $I$ and $Q$ branches in the analog parts of the RX can be modeled~as a quadrature mixer with an imbalanced LO signal given by~\cite{A:Advanced_methods_IQI_comp}
\begin{align}
\cos(\omega_\text{LO}t) - j\epsilon\sin(\omega_\text{LO}t+\phi) =%
K_1 e^{-j\omega_\text{LO}t} + K_2 e^{j\omega_\text{LO}t},\label{Eq:imbalance_LO}
\intertext{where the IQI coefficients $K_1$ and $K_2$ are given by}
K_1 = \frac{1}{2}\left(1+\epsilon e^{-j\phi}\right)\quad\text{and}\quad
K_2 = \frac{1}{2}\left(1-\epsilon e^{j\phi}\right).\label{Eq:K1&K2}
\end{align}

From \eqref{Eq:K1&K2} it is easy to see that $K_{1}$ and $K_{2}$ are connected through the relation
\begin{align}
K_1=1-K_2^\ast,
\label{Eq:K1connectK2}
\end{align}
while they can be used to define the $\rm{IRR}$~as
\begin{align}
\mathrm{IRR}=\frac{\abs{K_1}^2}{\abs{K_2}^2}.\label{Eq:IRR_tr}
\end{align}
It is noted that the $\rm{IRR}$ is a measure for the attenuation of the image frequency band. For practical analog RF front-end electronics, the $\rm{IRR}$ is typically in the range of $20-40$ dB \cite{A:Circularity_based_IQI_Comp_in_wideband_DC_RXs,
B:Ravazi,
A:Classical_and_modern_RX_archtectures,
A:DC_Radio_Transceivers_for_dig_com,
A:Effects_of_RF_impairments_In_Cascaded}, while for perfect matching, i.e., $\epsilon=1$ and $\phi=0$, we have $K_1=1$ and $K_2=0$ (i.e., $\rm{IRR}\to\infty$).

\begin{figure}
\psfrag{f}[t][][1]{$f$}%
\psfrag{k}[t][][1]{$k$}%
\psfrag{-2}[][l][0.8]{$-2$}%
\psfrag{-1}[][l][0.8]{$-1$}%
\psfrag{2}[][][0.8]{$2$}%
\psfrag{1}[][][0.8]{$1$}%
\psfrag{Baseband}[][][1]{\quad Baseband}%
\psfrag{RF signal}[][][1]{\quad RF signal}%
\psfrag{(a)}[][][1]{(a)}%
\psfrag{(b)}[][][1]{(b)}%
\psfrag{(c)}[][][1]{(c)}%
\centering\includegraphics[width=0.5\linewidth,trim=130 0 130 0,clip=false]{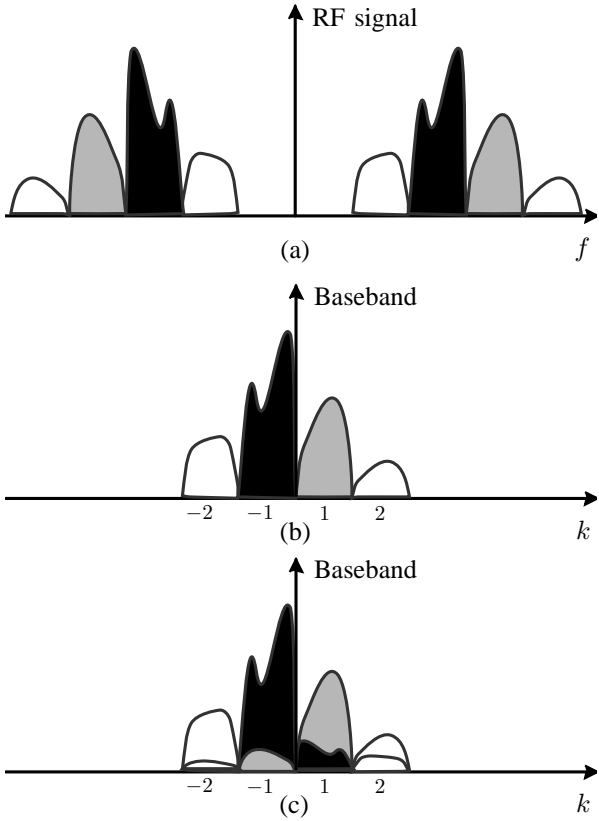}
\caption{Spectra of the (noise-free) received signals: (a)~before down-conversion (passband RF signal), (b)~after down-conversion, when ideal RF front-end is considered, and (c)~after down-conversion, when the RX RF front-end suffers from IQI, where for the $1$st subcarrier, the intermixing of the image signal (in black) and the desired signal (in grey) is clearly visible, while for the $-1$st subcarrier, the image and desired signals are depicted in grey and black colors, respectively.}
\label{fig:IQI_effects}
\end{figure}

Assuming that the I/Q mixer is the only source of imbalance, the resulting down-converted I/Q signal appears as
\begin{align}
g_\text{IQI}(t) = K_1 g(t) + K_2 g^\ast(t),\label{Eq:IQI_basic}
\end{align}
where $g(t)$ denotes the signal in the time domain under perfect I/Q balance.
The term $K_2 g^\ast(t)$ in \eqref{Eq:IQI_basic}, which is caused by the imbalance, represents the self-interference effect and results in crosstalk between the mirror frequencies in the down-converted signal.\footnote{In the context of this paper, we use the terms ``mirror-interference" and ``self-interference" to refer to the effects of IQI in the frequency and time domains, respectively.}\:This is illustrated in Fig.~\ref{fig:IQI_effects}, where subcarrier $k$ experiences interference from the signal received on the mirror subcarrier $-k$ and vice versa. Since complex conjugation in the time domain corresponds to complex conjugation and mirroring in the frequency domain, the spectrum of the imbalanced signal on subcarrier $k$ becomes
\begin{align}
G_\text{IQI}(k) = K_1 G(k) + K_2 G^\ast(-k),\label{Eq:IQI_basic_k}
\end{align}
where $G(k)$ and $G(-k)$ denote the spectra of the signal with perfect I/Q balance on subcarriers $k$ and $-k$, respectively.

\subsection{RX with IQI in the RF Front-end}\label{SS:SM_RX}

In this section, we explore the effect of I/Q mismatch on the overall link quality, assuming that the RF front-end of the TX is perfect, while the RX suffers from IQI.\footnote{This might correspond to a downlink scenario, where the low-cost DCA user equipment (UE) suffers from IQI.}\:We also assume that transmitted signals $s(k)$ and $s(-k)$, carried by subcarriers $k$ and $-k$, are associated with channel gains $h(k)$ and $h(-k)$, respectively, which are mutually independent RVs.

According to \eqref{Eq:IQI_basic_k}, the baseband equivalent received signal on subcarrier $k$ is given~by
\begin{align}
r(k) = K_1 r_\text{id}(k) + K_2 r_\text{id}^\ast(-k)\label{Eq:RX_IQI_1}
\end{align}
with $r_\text{id}(-k)$ being the baseband equivalent received signal for perfect I/Q balance on subcarrier $-k$.
Substituting \eqref{Eq:r_ideal} into \eqref{Eq:RX_IQI_1}, we obtain
\begin{align}
r(k) = K_1 h(k) s(k) + i(k) + w(k),\label{Eq:r_MC_RX_IQI}
\end{align}
where the interference and composite noise terms, namely $i(k)$ and $w(k)$, are given~by
\begin{align}
&i(k) = K_2 h^\ast(-k) s^\ast(-k),\label{Eq:i_k}\\
&w(k) = K_1 n(k) + K_2 n^\ast(-k).\label{Eq:w_k}
\end{align}
Here, $w(k)$ is a zero-mean complex Gaussian process with variance
\begin{align}
\sigma_w^2=\left(\abs{K_1}^2+\abs{K_2}^2\right) N_0.\label{Eq:var_w}
\end{align}

\section{The proposed IQSC Transceiver Design}\label{S:IQSC}

In this section, we present IQSC, a novel low-complexity scheme for increasing the performance of single-antenna multi-carrier communication systems suffering from IQI at the RX. The proposed architecture has two main components: a) the IQSC encoding scheme at the TX, and b) the combining scheme at the RX.

\subsection{The IQSC Encoding Scheme}\label{SS:IQSC_Enc}

Transmission is organized in two time intervals.
As shown in Table~\ref{T:Encodiong_and_transmission}, given the data set $S_K$, the following sequences of symbols are transmitted during the first and second time intervals, respectively,
\begin{align}
\mathrm{T}_1\left(S_K\right)={}&\left(s^\ast(-K),\dotsc,s^\ast(-k),\dotsc,s^\ast(-1),\right.\nonumber\\
&\left.s(1),\dotsc,s(k),\dotsc,s(K)\right),\label{Eq:T_1}\\
\mathrm{T}_2\left(S_K\right)={}&\left(s(K),\dotsc,s(k),\dotsc,s(1),\right.\nonumber\\
&\left.-s^\ast(-1),\dotsc,-s^\ast(-k),\dotsc,-s^\ast(-K)\right).\label{Eq:T_2}
\end{align}

\begin{table*}
\centering%
\renewcommand{\arraystretch}{1.3}
\caption{The IQSC encoding and transmission protocol.}
\label{T:Encodiong_and_transmission}
\begin{tabular}{|l||r|c|r|c|r|r|c|r|c|r|}
\hline
\textbf{Subcarrier index} & $-K$\phantom{)} & $\cdots$ & $-k$\phantom{)} & $\cdots$ & $-1$\phantom{)} & $1$\phantom{)} & $\cdots$ & $k$\phantom{)} & $\cdots$ & $K$\phantom{)} \\
\hline
\textbf{Intended data set} & $s\left(-K\right)$ & $\cdots$ & $s\left(-k\right)$ & $\cdots$ & $s\left(-1\right)$ & $s\left(1\right)$ & $\cdots$ & $s\left(k\right)$ & $\cdots$ & $s\left(K\right)$ \\
\hline\hline
\textbf{First time interval} & $s^{*}\left(-K\right)$ & $\cdots$ & $s^{*}\left(-k\right)$ &$\cdots$ & $s^{*}\left(-1\right)$ & $s\left(1\right)$ & $\cdots$ & $s\left(k\right)$ & $\cdots$ & $s\left(K\right)$ \\
\hline
\textbf{Second time interval} & $s\left(K\right)$ & $\cdots$ &$s\left(k\right)$ & $\cdots$ & $s\left(1\right)$ & $-s^{*}\left(-1\right)$ & $\cdots$ &$-s^{*}\left(-k\right)$&$\cdots$ &$-s^{*}\left(-K\right)$\\
\hline
\end{tabular}
\end{table*}

Assuming that the channels for subcarriers $k$ and $-k$ remain constant over two consecutive time intervals,\footnote{Practical systems are usually designed such that this assumptions holds in order to facilitate channel estimation and tracking. For channel estimation, the technique proposed in \cite{C:Div_Gain_TX_IQI} could be applied.} the received signal in the first time interval on subcarrier $k$ becomes
\begin{align}
x_1(k)={}&K_1 h(k)s(k) + K_2\left(h(-k)s^\ast(-k)\right)^\ast\nonumber\\
&+K_1 n_1(k) + K_2 n_1^\ast(-k)\nonumber\\
={}&a_1 s(k) + a_2 s(-k) + w_1(k),\label{Eq:x_syn_k_1}
\end{align}
where $a_1$ and $a_2$ are newly introduced channel-related parameters given~by
\begin{align}
a_1 = K_1 h(k) \quad\text{and}\quad a_2 = K_2 h^\ast(-k),\label{Eq:a1&a2}
\end{align}
and the composite noise term $w_1(k)$ in the first time interval on subcarrier $k$ is given by
\begin{align}
w_1(k) = K_1 n_1(k) + K_2 n_1^\ast(-k).\label{Eq:w_1_k}
\end{align}

Similarly, the received signal in the second time interval on subcarrier $k$~is
\begin{align}
x_2(k)={}&K_1h(k)\left(-s^\ast(-k)\right) + K_2 \left(h(-k) s(k)\right)^\ast\nonumber\\
&+K_1 n_2(k) + K_2 n_2^\ast(-k)\nonumber\\
={}&- a_1 s^\ast(-k) + a_2 s^\ast(k) + w_2(k),\label{Eq:x_syn_k_2}
\end{align}
with the composite noise term $w_2(k)$ in the second time interval on subcarrier $k$~being
\begin{align}
w_2(k) = K_1 n_2(k) + K_2 n_2^\ast(-k).\label{Eq:w_2_k}
\end{align}

The received signals in the first and second time intervals on subcarrier $-k$~are
\begin{align}
x_3(-k)={}&K_1 h(-k) s^\ast(-k) + K_2 \left(h(k) s(k)\right)^\ast\nonumber\\
&+K_1 n_1(-k) + K_2 n_1^\ast(k)\nonumber\\
={}&a_3 s^\ast(-k) + a_4 s^\ast(k) + w_1(-k),\label{Eq:x_plyn_k_1}
\end{align}
and
\begin{align}
x_4(-k)={}&K_1 h(-k) s(k) + K_2 \left(-h(k) s^\ast(-k)\right)^\ast\nonumber\\
&+K_1 n_2(-k) + K_2 n_2^\ast(k)\nonumber\\
={}&a_3 s(k) - a_4 s(-k) + w_2(-k),\label{Eq:x_plyn_k_2}
\end{align}
respectively, where the channel-related parameters $a_3$ and $a_4$ are given~by
\begin{align}
a_3 = K_1 h(-k)\quad\text{and}\quad a_4 = K_2 h^\ast(k),\label{Eq:a3&a4}
\end{align}
and $w_1(-k)$ and $w_2(-k)$ are the composite noise terms in the first and second time intervals on subcarrier $-k$, calculated from \eqref{Eq:w_1_k} and \eqref{Eq:w_2_k}, respectively.

From \eqref{Eq:x_syn_k_1}, \eqref{Eq:x_syn_k_2}, \eqref{Eq:x_plyn_k_1}, and \eqref{Eq:x_plyn_k_2}, we observe that the received signals on subcarriers $k$ and $-k$ are interfered by the received signals on subcarriers $-k$ and $k$, respectively.

\subsection{The IQSC Combining Scheme}\label{SS:IQSC_Comb}

The IQSC combiner uses the four received signals in the first and second time interval on subcarriers $k$ and $-k$ to form the following two signals
\begin{align}
y(k)&= a_1^\ast x_1(k) + a_2 x_2^\ast(k) + a_3^\ast x_4(-k) + a_4 x_3^\ast(-k),\label{Eq:y_k_comb}\\
y(-k)&= -a_1 x_2^\ast(k) + a_2^\ast x_1(k) + a_3 x_3^\ast(-k) - a_4^\ast x_4(-k),\label{Eq:y_Minus_k_comb}
\end{align}
which are subsequently used for decoding. Substituting \eqref{Eq:x_syn_k_1}, \eqref{Eq:x_syn_k_2}, \eqref{Eq:x_plyn_k_1}, and \eqref{Eq:x_plyn_k_2} into \eqref{Eq:y_k_comb} and \eqref{Eq:y_Minus_k_comb}, we obtain
\begin{align}
y(k)&= \sum_{i=1}^4 \abs{a_i}^2 s(k) + z(k),\label{Eq:y_comb_k_final}\\
y(-k)&= \sum_{i=1}^4 \abs{a_i}^2 s(-k) + z(-k),\label{Eq:y_comb_plyn_k_final}
\end{align}
where
\begin{align}
\sum_{i=1}^4 \abs{a_i}^2=\left(\abs{K_1}^2+\abs{K_2}^2\right) \left(\abs{h(k)}^2+\abs{h(-k)}^2\right)
\label{Eq:gain_IQSC}
\end{align}
is the gain achieved by IQSC, and
\begin{align}
z(k) &= a_1^\ast w_1(k) + a_2 w_2^\ast(k) + a_3^\ast w_2(-k) + a_4 w_1^\ast(-k),\label{Eq:z_k}\\
z(-k)&= - a_1 w_2^\ast(k) + a_2^\ast w_1(k) + a_3 w_1^\ast(-k) - a_4^\ast w_2(-k),\label{Eq:z_-k}
\end{align}
represent the noise components at the output of the combiner, which both have variance
\begin{align}
\sigma_z^2= \left(\abs{K_1}^2 + \abs{K_2}^2\right)^2\left(\abs{h(k)}^2+\abs{h(-k)}^2\right) N_0.\label{Eq:sigma_z}
\end{align}
\begin{figure*}
\psfrag{a1}[][][1]{$a_1$}%
\psfrag{a2}[][][1]{$a_2$}%
\psfrag{a3}[][][1]{$a_3$}%
\psfrag{a4}[][][1]{$a_4$}%
\centering\includegraphics[width=0.9\linewidth,trim=0 0 0 0,clip=false]{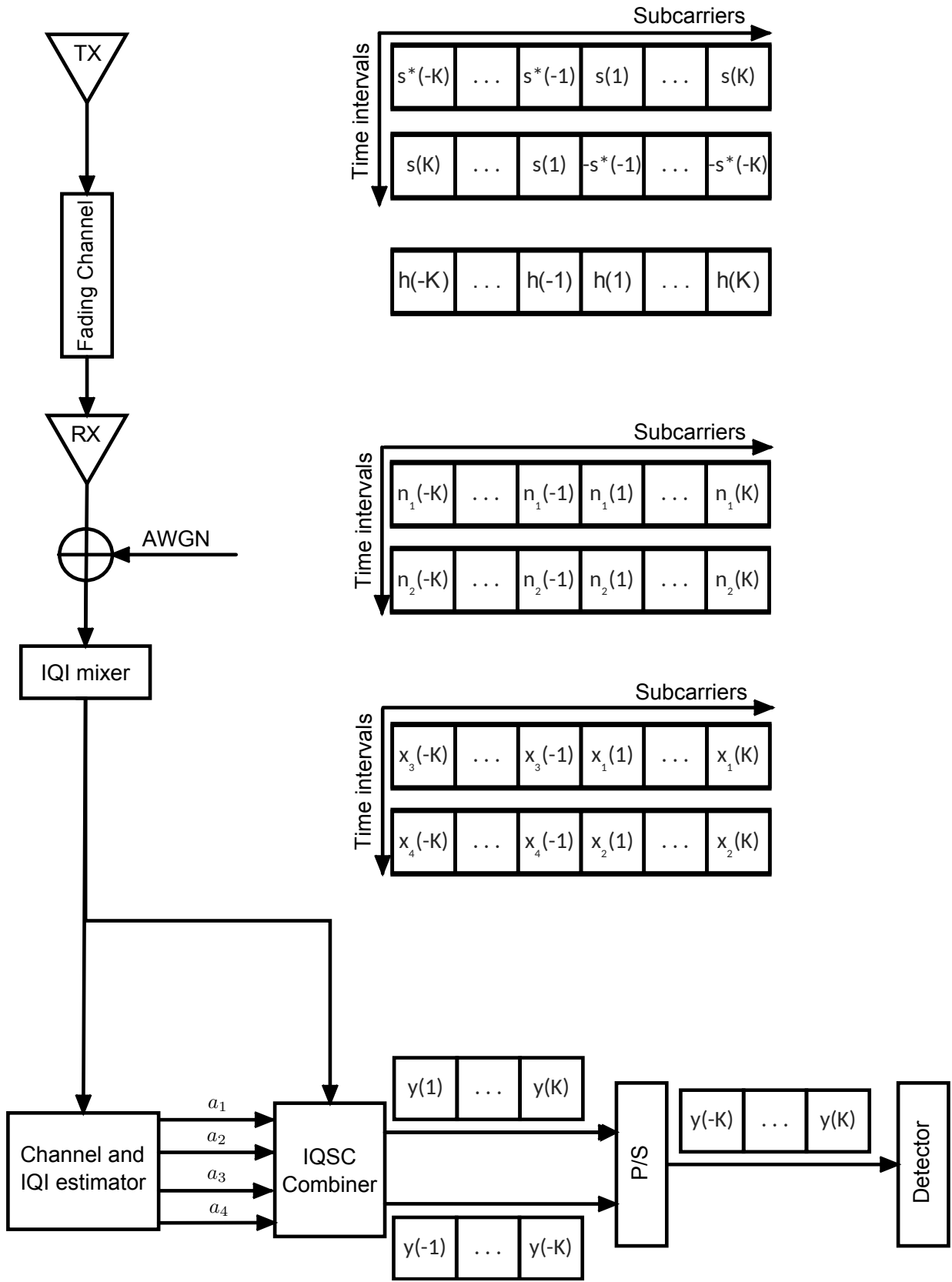}
\caption{Block diagram of the proposed IQSC scheme with RX IQI (P/S stands for the parallel-to-serial converter).}\label{fig:IQSC}
\end{figure*}

The combined signals in \eqref{Eq:y_comb_k_final} and \eqref{Eq:y_comb_plyn_k_final} are not only IQI free, but, at the same time, they reveal that the diversity order achieved by IQSC is equal to those of MRC with two antennas at the RX and Alamouti's space-time block code with two antennas at the TX~\cite{A:Alamouti}.

\subsection{An Alternative Mirror-Frequency Diversity Scheme}\label{SS:Alter}

Next, we present an alternative low-complexity transmit diversity scheme, i.e., A-IQSC, which is also applicable when the fading channel is constant only over a single time interval. Like IQSC, A-IQSC requires neither extra transmit power nor feedback. In fact, A-IQSC has the same properties, complexity, and rate as IQSC, since it uses two mirror subcarriers to transmit the same signal.

Particularly, given the intended data set $\tilde{S}_K$,
\begin{align}
\tilde{S}_K=\{s(1),\cdots,s(K)\}=\{s(k)\}_{k=1}^{K},
\end{align}
the following sequence of symbols is transmitted in one time interval
\begin{align}
\mathrm{T}(\tilde{S}_K) ={}&\left(s^\ast(K),\dotsc,s^\ast(k),\dotsc,s^\ast(1),\right.\nonumber\\
&\left.s(1),\dotsc,s(k),\dotsc,s(K)\right).\label{Eq:T_alter}
\end{align}

Then, the received signal at subcarriers $k$ and $-k$ are
\begin{align}
x(k)={}&K_1 h(k)s(k) + K_2\left(h(-k)s^\ast(k)\right)^\ast\nonumber\\
&+K_1 n(k) + K_2 n^\ast(-k)\nonumber\\
={}&\alpha s(k) + w(k),\label{Eq:Altenative_r_plus_k}
\end{align}
and
\begin{align}
x(-k)={}&K_1 h(-k)s^\ast(k) + K_2\left(h(k)s(k)\right)^\ast\nonumber\\
&+K_1 n(-k) + K_2 n^\ast(k)\nonumber\\
={}&\beta s^\ast(k) + w(-k),\label{Eq:Altenative_r_plyn_k}
\end{align}
respectively, where the new channel-related parameters are now given~by
\begin{align}
\alpha = a_1 + a_2 = K_1 h(k) + K_2 h^\ast(-k)\label{Eq:Alternative_alpha}
\end{align}
and
\begin{align}
\beta = a_3 + a_4 = K_1 h(-k) + K_2 h^\ast(k).\label{Eq:Alternative_beta}
\end{align}

The combiner, by using the received signals on subcarriers $k$ and $-k$, forms the following signal that is sent to the detector
\begin{align}
y(k)= \alpha^\ast x(k) + \beta x^\ast(-k).\label{Eq:Alternative_combiner}
\end{align}
By substituting \eqref{Eq:Altenative_r_plus_k} and \eqref{Eq:Altenative_r_plyn_k} into \eqref{Eq:Alternative_combiner}, we obtain
\begin{align}
y(k)= \left(\abs{\alpha}^2+\abs{\beta}^2\right) s(k) + z_c(k),\label{Eq:Alternative_combiner_final}
\end{align}
where
\begin{align}
z_c(k) = \alpha^\ast w(k) + \beta w^\ast(-k)\label{Eq:z_c_k}
\end{align}
is the noise component at the output of the combiner and has variance
\begin{align}
\sigma_{z_c}^2=\left(\abs{K_1}^2 + \abs{K_2}^2\right)^2\left(\abs{h(k)}^2+\abs{h(-k)}^2\right) N_0. \label{Eq:Alternative_noise_var}
\end{align}
Notice that the resulting combined signal in \eqref{Eq:Alternative_combiner_final} is IQI free. Furthermore, according to \eqref{Eq:Alternative_alpha} and \eqref{Eq:Alternative_beta}, and after some basic algebraic manipulations, we obtain
\begin{align}
\abs{\alpha}^2+\abs{\beta}^2={}&
\left(\abs{K_1}^2 + \abs{K_2}^2\right)\left(\abs{h(k)}^2+\abs{h(-k)}^2\right)\nonumber\\
&+4\re{K_1 K_2^\ast h(k) h(-k)}.\label{Eq:gain_alternative}
\end{align}
For practical values of $\phi$, i.e., $\phi<5^o$ \cite{A:OFDM_OR_IQI}, according to \eqref{Eq:K1&K2}, $K_1$ and $K_2$ can be approximated as $K_1\approx\re{K_1}$ and $K_2\approx\re{K_2}$, which based on \eqref{Eq:K1connectK2} are connected through
\begin{align}
\re{K_1} + \re{K_2^*}\approx 1,
\end{align}
or equivalently
\begin{align}
\frac{\re{K_2^\ast}}{\re{K_1}}\approx \frac{1}{\re{K_1}}-1.
\label{K2_K1_approx_con}
\end{align}
Moreover, for practical values of $\rm{IRR}$, i.e., $20\text{ }\rm{dB}\leq\rm{IRR}\leq 40\text{ }\rm{dB}$ \cite{A:Circularity_based_IQI_Comp_in_wideband_DC_RXs,B:Ravazi,A:Classical_and_modern_RX_archtectures,
A:DC_Radio_Transceivers_for_dig_com}, we have $\frac{1}{10}\leq\abs{\frac{\re{K_2^\ast}}{\re{K_1}}}\leq \frac{1}{100}$. By substituting these values into \eqref{K2_K1_approx_con}, we obtain $\frac{10}{11}\leq \re{K_1}\leq \frac{100}{101}$.
Hence, as $\rm{IRR}$ increases, $K_1\to 1$ and consequently
\begin{align}
K_1K_2^\ast = K_1 - K_1^2\to 0.
\end{align}
Furthermore, except for few central subcarriers, the correlation between subcarrier $k$ and its image subcarrier $-k$ is small due to their large spectral separation, hence, they can be assumed to be independent, i.e., $E\left\{4\re{K_1 K_2^\ast h(k) h(-k)}\right\}=0$.
Therefore, the second term on the right-hand side of \eqref{Eq:gain_alternative} can be neglected, and \eqref{Eq:gain_alternative} can be approximated as~\cite{A:OFDM_OR_IQI}
\begin{align}
\abs{\alpha}^2+\abs{\beta}^2\approx\left(\abs{K_1}^2 + \abs{K_2}^2\right) \left(\abs{h(k)}^2 + \abs{h(-k)}^2\right),
\label{Eq:gain_alternative_approx}
\end{align}
which is identical to the combined channel gain for IQSC in \eqref{Eq:gain_IQSC}. Since IQSC and A-IQSC also exhibit the same noise variance (see \eqref{Eq:sigma_z} and \eqref{Eq:Alternative_noise_var}), we expect A-IQSC to achieve a similar performance as IQSC.

\section{Performance analysis}\label{S:Perf}

In this section, we investigate the effects of IQI on the system performance considering the cases of 1) ideal RF front-end, 2) uncompensated IQI at the RX (i.e., IQI RF front-end), 3) compensated IQI using IQSC, and 4) compensated IQI using A-IQSC. Thereby, perfect CSI and IQI parameter knowledge is assumed at the RX.\footnote{The case of imperfect CSI due to IQI and outdated estimation will not be investigated in this paper due to space limitations. However, the performance results presented here can be considered as upper bounds for the case of imperfect CSI.}\:In Section~\ref{SS:Perf_SINR}, we give the respective instantaneous $\rm{SINR}$ expressions, and then, in Section~\ref{SS:Perf_OP}, we use these expressions to evaluate the end-to-end outage probability, i.e., the probability that the SINR falls below a given threshold. In Section~\ref{SS:Perf_SER}, we derive closed-form expressions for the SER.

\subsection{Signal-to-Interference-plus-Noise Ratio}\label{SS:Perf_SINR}

\subsubsection{Ideal RF front-end}\label{SSS:Perf_SINR_Ideal}

In case of an ideal RF front-end, the instantaneous SINR is given by
\begin{align}
\gamma_\text{id}(k) = \abs{h(k)}^2\frac{E_s}{N_0},\label{Eq:gamma_ideal}
\end{align}
where $E_s$ is the average energy of the transmitted symbol.

\subsubsection{IQI RF front-end}\label{SSS:Perf_SINR_RX}

When IQI impairs the RX, then, based on \eqref{Eq:r_MC_RX_IQI} and \eqref{Eq:w_k}, the instantaneous SINR per symbol on subcarrier $k$ is given~by
\begin{align}
\gamma(k)={}&\frac{\abs{K_1}^2\abs{h(k)}^2 E_s}{\abs{K_2}^2\abs{h(-k)}^2 E_s+\left(\abs{K_1}^2+ \abs{K_2}^2\right)N_0}\nonumber\\
={}&\frac{\gamma_\text{id}(k)}{\frac{\gamma_\text{id}(-k)}{\mathrm{IRR}}+\left(1+\frac{1}{\mathrm{IRR}}\right)}.
\label{Eq:gamma_MC_RX}
\end{align}

We assume that the correlation between subcarrier $k$ and its image $-k$ is small due to their large spectral separation.
Hence, $\gamma_\text{id}(k)$ and $\gamma_\text{id}(-k)$ can be assumed statistically independent.
In general, $\gamma_\text{id}(k)$ and $\gamma_\text{id}(-k)$ are correlated RVs with the correlation coefficient given by $\rho=\av{\gamma_\text{id}(k)\gamma_\text{id}^\ast(-k)}$.
When subcarriers $k$ and $-k$ are close to each other, then the correlation may be significant. To simplify the analysis, we assume that $\rho=0$, which is an accurate assumption except for few central subcarriers \cite{A:IQI_SINR_ODFM,A:IQI_in_AF_Nakagami_m,A:OFDM_OR_IQI}. Note that this assumption is valid for several practical wireless communication systems, such as LTE, high performance radio local area network (HIPERLAN), and WLAN, since the central subcarriers are not used in those systems~\cite{B:4G_LTE,A:Wifi_unused_DC_subcarriers,A:Virtual_Carriers,A:Virtual_Carriers_Tallambura}.

\subsubsection{IQSC}\label{SSS:Perf_SINR_IQSC}

Assuming equal transmit power for the IQSC scheme and the IQI RF front-end scenario, then, based on \eqref{Eq:y_comb_k_final}, \eqref{Eq:y_comb_plyn_k_final}, and \eqref{Eq:sigma_z}, the instantaneous $\rm{SNR}$ is given~by
\begin{align}
\gamma_\text{IQSC}(k)={}&
\left(\abs{h(k)}^2+\abs{h(-k)}^2\right)\frac{E_s}{2N_0}\nonumber\\
={}&\frac{1}{2}\left(\gamma_\text{id}(k) + \gamma_\text{id}(-k)\right).\label{Eq:gamma_P}
\end{align}

\subsubsection{A-IQSC}\label{SSS:Perf_SINR_Alter}

Assuming that the total transmit power of both subcarriers in A-IQSC is identical to that of the IQI RF front-end scenario, then, based on \eqref{Eq:Alternative_combiner_final}, \eqref{Eq:Alternative_noise_var}, and \eqref{Eq:gain_alternative_approx}, for practical IQI levels, the instantaneous $\rm{SNR}$ can be approximated by~\eqref{Eq:gamma_P}.

\subsection{Outage Probability Analysis}\label{SS:Perf_OP}

\subsubsection{Ideal RF front-end}\label{SSS:Perf_OP_Ideal}

Assuming that the channel amplitude $\abs{h(k)}$ follows a Rayleigh distribution, the instantaneous SNR $\gamma_\text{id}(k)$ given by \eqref{Eq:gamma_ideal} is an exponential distributed RV. Hence, the end-to-end outage probability is given~by
\begin{align}
P_\text{out}\left(\gamma_\text{th}\right)=1-e^{-\frac{\gamma_\text{th}}{\overline{\gamma}_\text{id}}},\label{Eq:P_out_ideal}
\end{align}
where $\overline{\gamma}_\text{id}=E_s/N_0$, and $\gamma_\text{th}=2^R-1$ is the SNR threshold with $R$ being the transmission rate.

\subsubsection{IQI RF front-end}\label{SSS:Perf_OP_RX}

Taking into account \eqref{Eq:gamma_MC_RX}, using $x=\gamma_\text{id}(k)$ and $y=\gamma_\text{id}(-k)$, and exploiting the independence between the exponentially distributed RVs $x$ and $y$, we obtain for the outage probability~\cite{A:OFDM_OR_IQI,B:Probability_book_stark}
\begin{align}
P_\text{out}\left(\gamma_\text{th}\right)= \int_0^\infty F\left(x<\frac{\gamma_\text{th}}{\mathrm{IRR}}y + \gamma_\text{th}\left(1+\frac{1}{\mathrm{IRR}}\right)\right)f(y)dy,\label{Eq:Pout_RX_start}
\end{align}
where $F(x<X)$ and $f(y)$ are the cumulative distribution function (CDF) of $x$ and the probability density function (PDF) of $y$, respectively.
Evaluating the integral in \eqref{Eq:Pout_RX_start}, we get
\begin{align}
P_\text{out}\left(\gamma_\text{th}\right)= 1-\frac{e^{-\frac{\gamma_\text{th}}{\overline{\gamma}_\text{id}}\left(1+\frac{1}{\mathrm{IRR}}\right)}}
{1+\frac{\gamma_\text{th}}{\mathrm{IRR}}},\label{Eq:Pout_RX_final}
\end{align}
which depends in the IQI via the $\rm{IRR}$. Note that, in the high-$\rm{SNR}$ regime ($\overline{\gamma}_\text{id}\to\infty$), the outage probability~approaches
\begin{align}
P_\text{out}\left(\gamma_\text{th}\right)\circeq 1 - \frac{1}{1+\frac{\gamma_\text{th}}{\mathrm{IRR}}}.\label{Eq:Outage_Floor}
\end{align}
Furthermore, notice that in case of an ideal RF front-end, i.e., $\rm{IRR}\to\infty$, \eqref{Eq:Pout_RX_final} simplifies to~\eqref{Eq:P_out_ideal}.

\subsubsection{IQSC}\label{SSS:Perf_OP_IQSC}

In the proposed scheme, the effective $\rm{SNR}$, $\gamma_\text{IQSC}$, is the sum of the instantaneous $\rm{SNR}s$ on subcarriers $k$ and $-k$, which follow exponential distributions. Therefore, $\gamma_\text{IQSC}$ is a chi-square distributed RV with mean $\overline{\gamma}_{\text{IQSC}}$ and the PDF is given~by
\begin{align}
f_{\gamma_\text{IQSC}}(\gamma)= \frac{\gamma}{\overline{\gamma}_\text{IQSC}^2}
e^{-\frac{\gamma}{\overline{\gamma}_\text{IQSC}}}.\label{Eq:PDF_IQSC}
\end{align}
Hence, the outage probability is obtained~as
\begin{align}
P_\text{out}\left(\tilde{\gamma}_\text{th}\right)= 1-\frac{\overline{\gamma}_\text{IQSC} + \tilde{\gamma}_\text{th}}{\overline{\gamma}_\text{IQSC}}
e^{-\frac{\tilde{\gamma}_\text{th}}{\overline{\gamma}_\text{IQSC}}},\label{Eq:P_out_IQSC}
\end{align}
where $\overline{\gamma}_\text{IQSC}=\overline{\gamma}_{\text{id}}$, and $\tilde{\gamma}_{th}$ is the $\rm{SNR}$ threshold. Since, IQSC requires two time slots to transmit one data set $S_K$, $\tilde{\gamma}_{th}$ and $R$ are connected via $\tilde{\gamma}_\text{th} = 2^{2R}-1.$ Interestingly, when IQSC is employed, the outage performance is independent of the levels of IQI at the RX, i.e., it is not a function of $\rm{IRR}$.

\subsubsection{A-IQSC}\label{SSS:Perf_OP_Alter}

In A-IQSC and for practical IQI levels, according to \eqref{Eq:gain_alternative_approx}, the effective $\rm{SNR}$ is approximately equal to $\gamma_\text{IQSC}$. Hence, the outage probability of A-IQSC is approximately equal to that of IQSC.

\subsection{Symbol Error Rate Analysis}\label{SS:Perf_SER}

For slow flat fading, the SER can be derived by averaging the conditional error probability in AWGN, $P_s(e|\gamma)$, over the fading distribution. Mathematically, the SER can be evaluated~as
\begin{align}
P_s(e)=\int_0^\infty P_s(e|\gamma)f_\gamma(\gamma)d\gamma,\label{Eq:SER_general}
\end{align}
where $f_\gamma(\gamma)$ is the PDF of the end-to-end SNR. For several Gray bit-mapped constellations employed in practical systems, $P_s(e|\gamma)$ is of the form
\begin{align}
P_s(e|\gamma)= A\erfc\left(\sqrt{B\gamma}\right),\label{Eq:SER_AWGN}
\end{align}
where $A$ and $B$ are modulation dependent constants.
For example, for binary phase-shift keying (BPSK) $A=0.5$ and $B=1$, while for quadrature phase-shift keying (QPSK) $A=1$ and $B=0.5$. In the case of square/rectangular $M$-quadrature amplitude modulation (QAM), $P_s(e|\gamma)$ can be written as a finite weighted sum of $\erfc\left(\sqrt{B\gamma}\right)$~\cite{B:Dig_Com_Fading_Channels}.

\subsubsection{Ideal RF front-end}\label{SSS:Perf_SER_Ideal}

The instantaneous $\rm{SNR}$ follows an exponential distribution with PDF
\begin{align}
f_{\gamma_\text{id}}(\gamma)= \frac{1}{\overline{\gamma}_\text{id}}
e^{-\frac{\gamma}{\overline{\gamma}_\text{id}}}.\label{Eq:PDF_exponential}
\end{align}
Substituting \eqref{Eq:SER_AWGN} and \eqref{Eq:PDF_exponential} into \eqref{Eq:SER_general}, and carrying out the integration, we get
\begin{align}
P_s(e)= \frac{A}{1+B\overline{\gamma}_\text{id}+\sqrt{B\overline{\gamma}_\text{id}+
(B\overline{\gamma}_\text{id})^2}}.\label{Eq:SER_ideal}
\end{align}

\subsubsection{IQI RF front-end}\label{SSS:Perf_SER_RX}

In case of uncompensated IQI, the SER is given by~\cite{C:SER_RX_IQI}
\begin{align}
P_s(e)= \frac{1}{M}\sum_{m=1}^M P_s(e_m),\label{Eq:SER_RX}
\end{align}
where $M$ is the modulation order and
\begin{align}
P_s(e_m)={}&\frac{M-1}{M}-\frac{2(\sqrt{M}-1)}{M\sqrt{1+\phi_m^2}}\nonumber\\
&-\frac{4(M-1)^2}{\pi M\sqrt{1+\phi_m^2}}\arctan\left(\frac{1}{1+\phi_m^2}\right)\label{Eq:SER_RX_m}
\end{align}
with
\begin{align}
\phi_m^2= \frac{2(M-1)}{3E_s\mathrm{IRR}}\left(\abs{i_m}^2+\left(\mathrm{IRR}+1\right)N_0\right).\label{Eq:phi_m}
\end{align}
In \eqref{Eq:phi_m}, $i_{m}$ denotes the $m$-th complex-valued symbol of the modulation alphabet. From \eqref{Eq:SER_RX_m} and \eqref{Eq:phi_m}, we observe that for uncompensated IQI, the $\rm{SER}$ depends on the levels of the IQI via $\rm{IRR}$.

\subsubsection{IQSC}\label{SSS:Perf_SER_IQSC}

By substituting \eqref{Eq:PDF_IQSC} and \eqref{Eq:SER_AWGN} into \eqref{Eq:SER_general} and carrying out the integration, we get after some basic algebraic manipulations
\begin{align}
P_s(e) = A\frac{-3\sqrt{B}-2\overline{\gamma}_\text{IQSC}\sqrt{B^3}+2(B+1)\sqrt{B+1/\overline{\gamma}_\text{IQSC}}}
{2\overline{\gamma}_\text{IQSC}\sqrt{\left(B+1/\overline{\gamma}_\text{IQSC}\right)^3}}.\label{Eq:SER_IQSC}
\end{align}
From \eqref{Eq:SER_IQSC}, we observe that by employing IQSC, the error performance of a receiver with IQI becomes independent of the IQI level.

\subsubsection{A-IQSC}\label{SSS:Perf_SER_Alter}

For practical values of $\rm{IRR}$, based on \eqref{Eq:gain_alternative_approx}, the SER can be straightforwardly approximated by \eqref{Eq:SER_IQSC}.

\subsection{Comparison with Equal-Rate Repetition Coding (RC)}\label{SS:Comp}

In this section, we compare the performance of the proposed IQSC with RC across subcarriers, where the same signal is sent from the TX to the RX twice using subcarriers $k$ and $-k$. We make this comparison in order to demonstrate the efficiency of IQSC compared to a scheme having the same rate. Besides, since IQSC and RC assume the same form in the absence of IQI, we expect then to achieve similar performance in this case.

Assuming a system with RX IQI, the received signals on subcarriers $k$ and $-k$ are
\begin{align}
x(k) = K_1 h(k) s(k) + K_2 h^\ast(-k) s^\ast(k) + w(k),\label{Eq:x_1_DD}
\end{align}
and
\begin{align}
x(-k) = K_1 h(-k) s(k) + K_2 h^\ast(k) s^\ast(k) + w(-k),\label{Eq:x_2_DD}
\end{align}
respectively, where we set $s(k)=s(-k)$ due to the RC.
In \eqref{Eq:x_1_DD} and \eqref{Eq:x_2_DD}, $w\left(k\right)$ and $w\left(-k\right)$ are given by \eqref{Eq:w_k}.

After combining the received signal using MRC, the overall signal can be written~as
\begin{align}
y_\text{RC}(k)&= h^\ast(k) x(k) + h^\ast(-k) x(-k)\nonumber\\
&=K_1\left(\abs{h(k)}^2+\abs{h(-k)}^2\right) s(k) + i_\text{RC}(k) + z_\text{RC}(k),\label{Eq:y_DD_IQI}
\end{align}
where $i_\text{RC}(k)$ and $z_\text{RC}(k)$ are the self-interference and noise terms, respectively, given~by
\begin{align}
i_\text{RC}(k)={}&2 K_2 h^\ast(k) h^\ast(-k) s^\ast(k),\label{Eq:i_D}\\
z_\text{RC}(k)={}&K_1 h^\ast(k) n(k) + K_2 h^\ast(k) n^\ast(-k)\nonumber\\
&+K_1 h^\ast(-k) n(-k) + K_2 h^\ast(k) n^\ast(k).\label{Eq:z_D}
\end{align}
Under perfect I/Q matching, i.e., $K_1=1$ and $K_2=0$, the model in \eqref{Eq:y_DD_IQI} reduces~to
\begin{align}
y_\text{RC}(k)= \left(\abs{h_1(k)}^2 + \abs{h_2(k)}^2\right) s(k) + w(k).\label{Eq:DD_SM_ideal}
\end{align}
In this case, since \eqref{Eq:DD_SM_ideal} is similar to \eqref{Eq:y_comb_k_final}, the outage probability can be written as in \eqref{Eq:P_out_IQSC}, and the SER is given by \eqref{Eq:SER_IQSC}, i.e., in the absence of IQI, IQSC and RC achieve indeed the same~performance.

Based on \eqref{Eq:y_DD_IQI}-\eqref{Eq:z_D}, the instantaneous $\rm{SINR}$ can be expressed~as
\begin{align}
\gamma_\text{RC}(k)= \frac{S_\text{RC}}{I_\text{RC}+N_\text{RC}},\label{Eq:SNR_DD_IQI}
\end{align}
where $S_\text{RC}$, $I_\text{RC}$, and $N_\text{RC}$ are the signal, interference, and noise power components, respectively, and are given~by
\begin{align}
S_\text{RC}&= \abs{K_1}^2\left(\abs{h(k)}^2 + \abs{h(-k)}^2\right)^2 E_s,\label{Eq:S_DD}\\
I_\text{RC}&= 4 \abs{K_2}^2 \abs{h(k)}^2\abs{h(-k)}^2 E_s,\label{Eq:I_DD}\\
N_\text{RC}&= \left(\abs{K_1}^2+\abs{K_2}^2\right)\left(\abs{h(k)}^2+\abs{h(-k)}^2\right) N_0.\label{Eq:N_DD}
\end{align}
Note that, in the high $\rm{SNR}$ regime, i.e., $N_0\to 0$, the instantaneous $\rm{SINR}$ can be obtained~as
\begin{align}
\gamma_\text{RC}(k)\circeq{}&\frac{\left(\abs{h(k)}^2+\abs{h(-k)}^2\right)^2}
{\abs{h(k)}^2\abs{h(-k)}^2}\frac{\mathrm{IRR}}{4}\nonumber\\
={}&\left(\frac{\abs{h(k)}^2}{\abs{h(-k)}^2}+\frac{\abs{h(-k)}^2}{\abs{h(k)}^2} + 2\right)\frac{\mathrm{IRR}}{4}.\label{Eq:SNR_DD_upper_limit_2}
\end{align}
Considering \eqref{Eq:SNR_DD_IQI} and \eqref{Eq:SNR_DD_upper_limit_2}, the effect of IQI on RC is fundamentally different from the effect of IQI on ordinary single time-slot (i.e., full-rate) systems. The interference term is not only affected by $h(-k)$ but also by $h(k)$. Unlike IQSC, although RC has the same rate, its $\rm{SINR}$ is upper bounded due to the IQI even for $N_{0}\to 0$, as described by \eqref{Eq:SNR_DD_upper_limit_2}. Hence, in the presence of IQI, IQSC achieves a better performance than RC.

\section{Numerical and Simulation Results}\label{S:Sim}

In this section, the performance of the schemes proposed in Section~\ref{S:IQSC} is illustrated in terms of outage and error probabilities for linearly modulated signals. In all cases, we validate the theoretical results (indicated with lines in the plots) with Monte-Carlo simulations (indicated with markers in the plots) so as to verify the accuracy of the closed-form expressions derived in Section~\ref{S:Perf}.
Exceptions to the rule are the A-IQSC, RC, and FTBC \cite{A:FTBC} schemes, where we (abusively) use lines with markers to plot simulation results. We compare the performance of the two proposed IQI coordination schemes (i.e., IQSC and A-IQSC) to that of the IQI RF front-end and the ideal RF front-end systems. The following assumptions are made for all schemes: a)~equal total transmit power, b)~uncorrelated channel gains for a subcarrier and its image, c)~equal average signal power on each subcarrier, and d)~perfect CSI and IQI parameter knowledge.

\begin{figure}
\centering\includegraphics[width=0.95\linewidth,trim=0 0 0 0,clip=false]{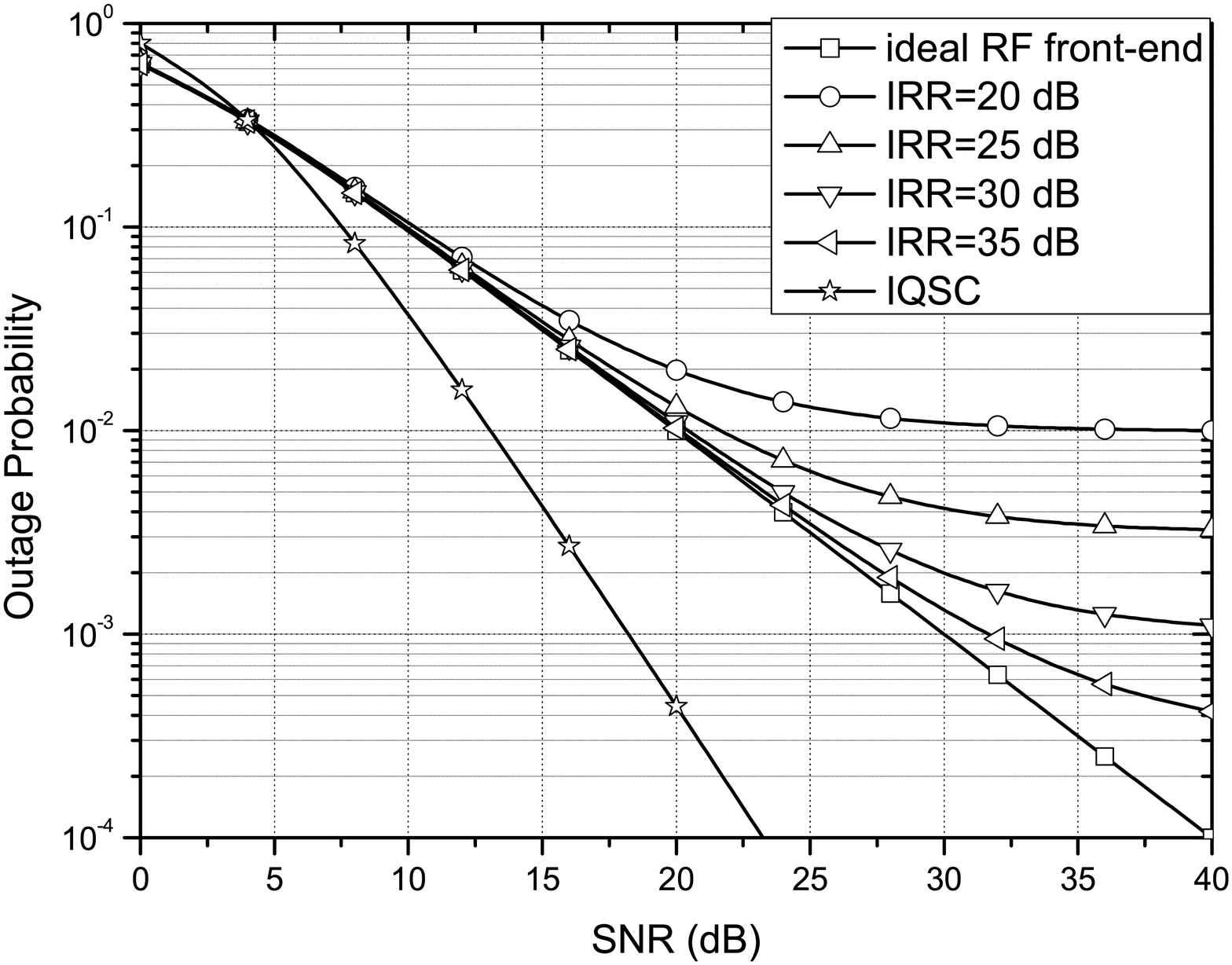}
\caption{Outage probability versus SNR for transmission rate $1$ bit/s/Hz and different IQI levels.}
\label{fig:outage_vs_SNR_1}
\end{figure}

\begin{figure}
\centering\includegraphics[width=0.95\linewidth,trim=0 0 0 0,clip=false]{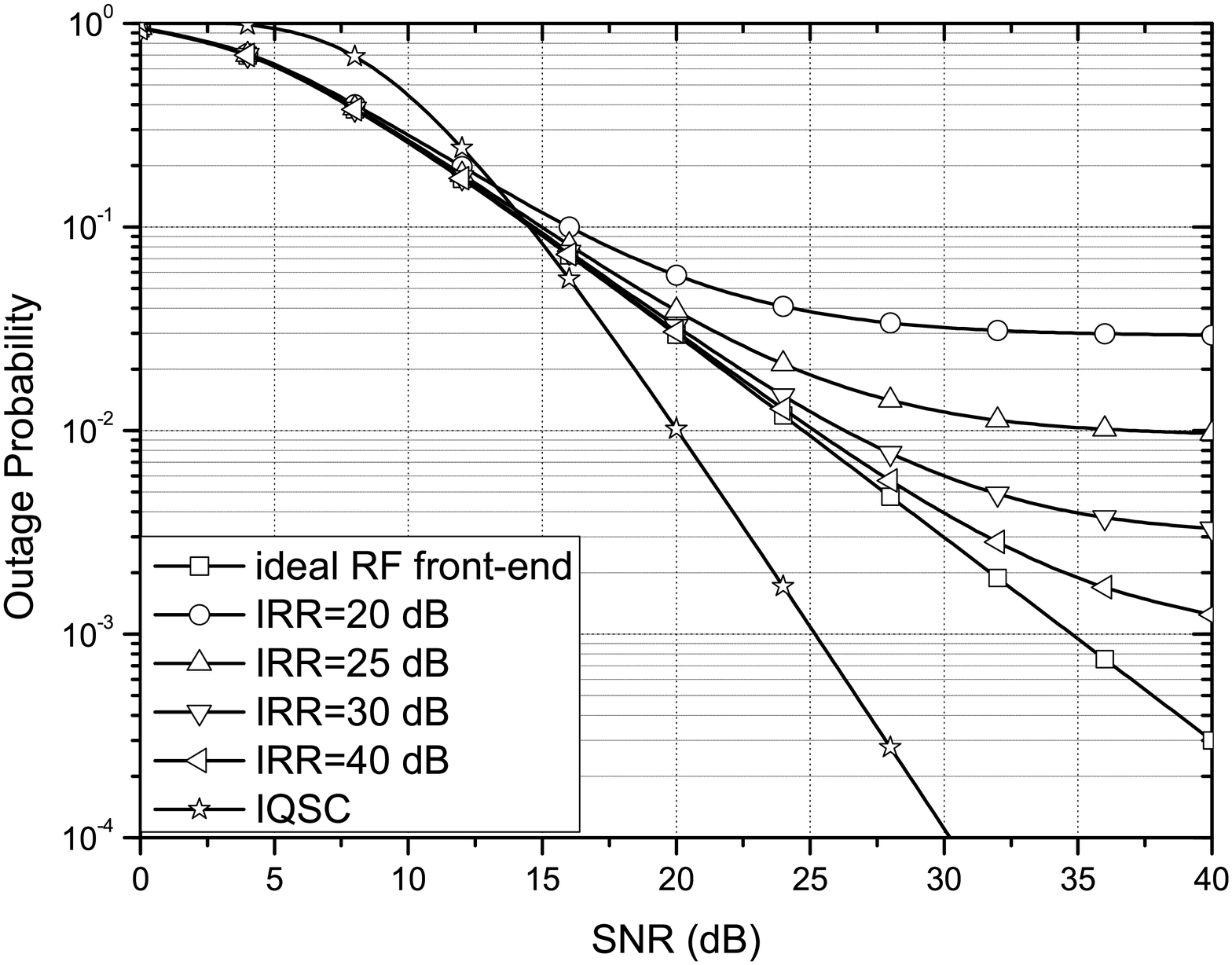}
\caption{Outage probability versus SNR for transmission rate $2$ bit/s/Hz and different IQI levels.}
\label{fig:outage_vs_SNR_2}
\end{figure}

The outage probability versus the $\rm{SNR}$ for the IQSC, IQI RF front-end, and ideal RF front-end systems is shown in Figs.~\ref{fig:outage_vs_SNR_1} and~\ref{fig:outage_vs_SNR_2} for transmission rates of $1$ bit/sec/Hz and $2$ bit/sec/Hz, respectively, and $\rm{IRR}$ values of $20,25,30$, and $35$ dB. We observe that the simulation results confirm the analytical expressions over the entire $\rm{SNR}$ range for all IQI levels. Interestingly, IQSC achieves a lower outage probability than the ideal RF front-end for practical availability requirements, and the corresponding gain increases for increasing $\rm{SNR}$ because of the achieved diversity. Indicatively, for transmission rates equal to $R=1$ bit/sec/Hz and $R=2$ bit/sec/Hz, IQSC outperforms the ideal RF front-end for $\rm{SNR}$ values greater than $4$ dB and outage probabilities lower than $0.08$, respectively.

\begin{figure}
\centering\includegraphics[width=0.95\linewidth,trim=0 0 0 0,clip=false]{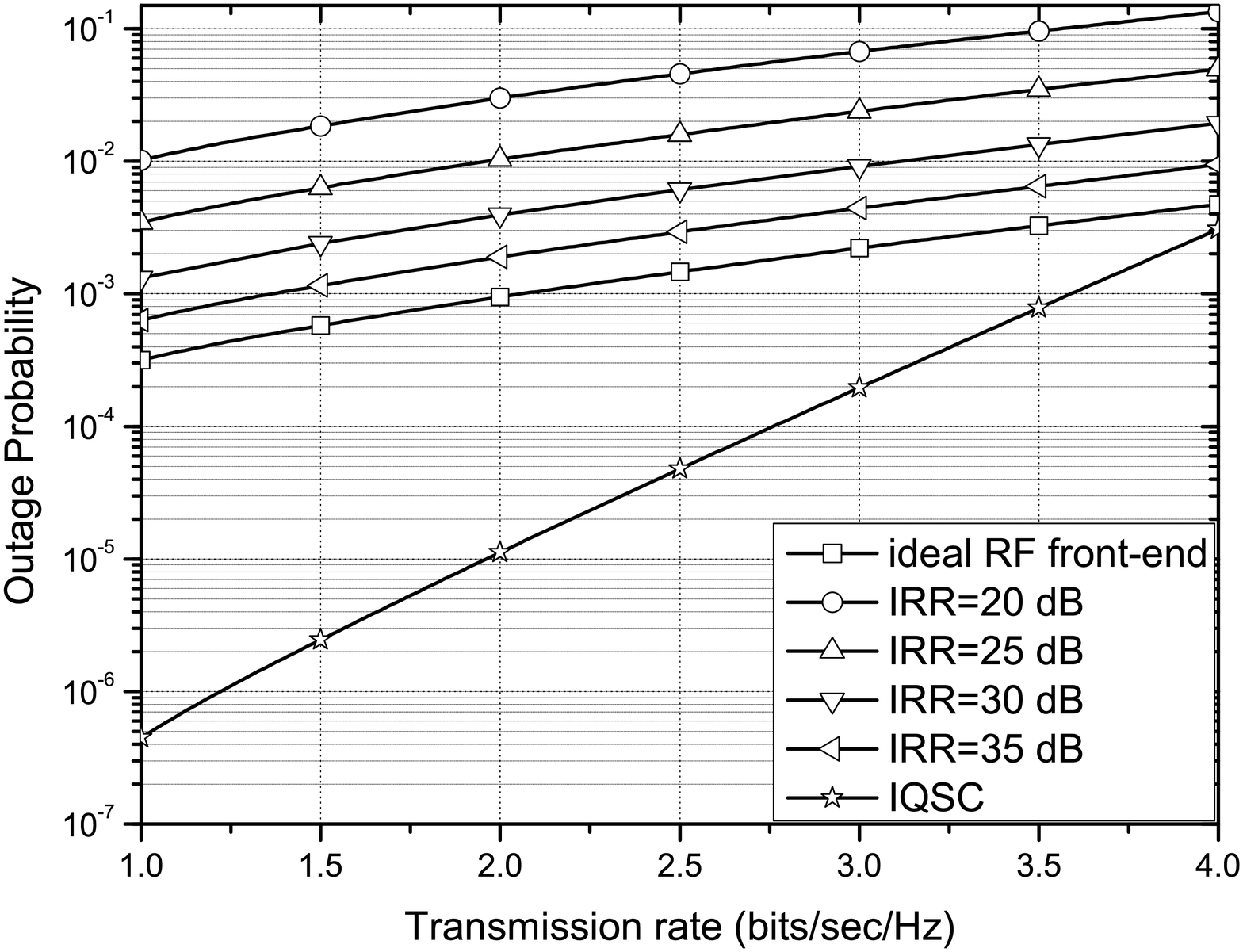}
\caption{Outage probability versus transmission rate for SNR equal to $35$ dB and different IQI levels.}
\label{fig:outage_vs_transmission_rate}
\end{figure}

In Fig.~\ref{fig:outage_vs_transmission_rate}, the outage probability is plotted as a function of transmission rate, $R$, for the IQSC, IQI RF front-end, and ideal RF front-end systems for an $\rm{SNR}$ of $35$ $\rm{dB}$ and $\rm{IRR}$ values of $20,25,30$, and $35$ $\rm{dB}$. It can be observed that, for practical values of $R$, IQSC significantly outperforms the ideal RF front-end. For example, for $R=2$ bit/s/Hz and $\mathrm{IRR}=20$ dB, the outage probability of IQSC is $99\%$ and $98.8\%$ lower than those of the IQI RF front-end and the ideal RF front-end systems, respectively, while for $R=4$ bit/s/Hz the corresponding values are $97.7\%$ and $34.9\%$.

\begin{figure}
\centering\includegraphics[width=0.95\linewidth,trim=0 0 0 0,clip=false]{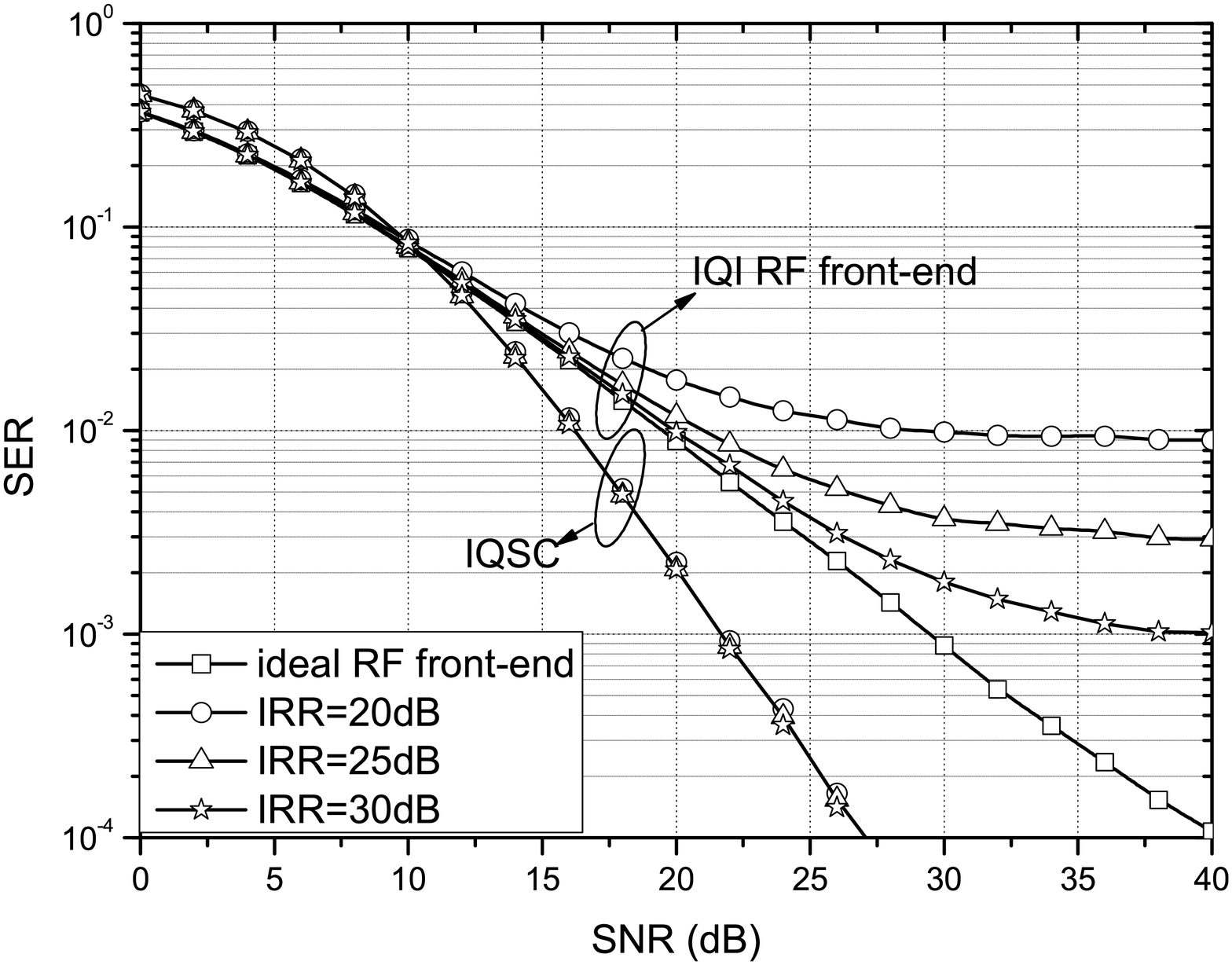}
\caption{SER versus SNR for different values of $\mathrm{IRR}$ and QPSK.}
\label{fig:QPSK_SER_2_IRR}
\end{figure}

To further illustrate the effect of IQI on the error performance, in Fig.~\ref{fig:QPSK_SER_2_IRR}, the average SER is plotted for QPSK and different values of $\rm{IRR}$. It can be observed that, without IQI compensation, the lower the $\rm{IRR}$, the larger the average SER becomes. Furthermore, as expected, the performance of IQSC is not affected by the values of the IQI parameters. This finding indicates that, for low-cost devices, IQSC can achieve a significant diversity gain without the use of additional antennas by coordinating the IQI effect through appropriate block coding and signal processing.

\begin{figure}
\centering\includegraphics[width=0.95\linewidth,trim=0 0 0 0,clip=false]{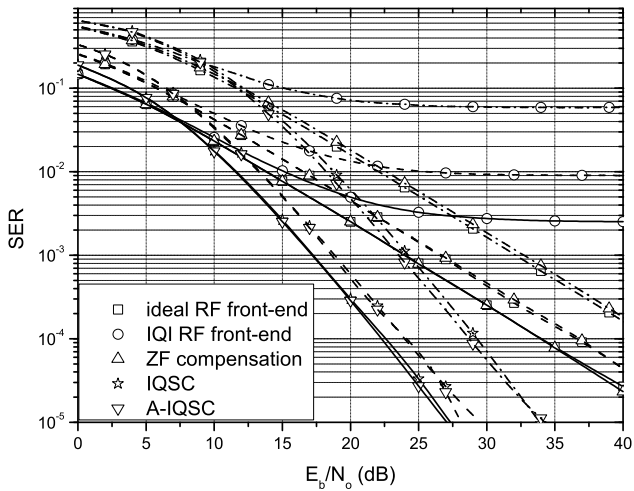}
\caption{SER versus SNR per bit for $\mathrm{IRR}=20$ dB, $\phi=5^{o}$, and BPSK (solid), QPSK (dashed), and $16$-QAM (dash-dotted).}
\label{fig:SER2BNR}
\end{figure}

To demonstrate the efficiency of IQSC for different modulations schemes \cite{A:RC_BB_Comp_TXRX_IQI_Mob_MIMO_OFDM,A:Broadband_MIMO_OFDM_WC,A:LFPC_MIMO_OFDM,B:Goldsmith,
B:proakis2007digital}, the SER versus the $\rm{SNR}$ per bit is plotted in Fig.~\ref{fig:SER2BNR} for BPSK, QPSK, and $16$-QAM, and $\mathrm{IRR}=20$ dB assuming the following five systems: i)~ideal RF front-end, ii)~IQI RF front-end, iii)~compensated IQI with zero forcing (ZF) \cite{A:DD_Channel_Est_and_high_IQI_Comp_in_OFDM_RXs,Mokhtar_IQI,A:RC_BB_Comp_TXRX_IQI_Mob_MIMO_OFDM}, iv)~IQSC, and v)~A-IQSC. As expected, the error performance of ZF is very close to the case of the ideal RF front-end.
This indicates that the latter scheme can be used as a benchmark for the evaluation of the IQSC performance.
Besides, as expected, IQSC and A-IQSC achieve almost the same SER. Interestingly, at a SER of $10^{-3}$, which is a practical requirement in several wireless communication standards \cite{TR:3GPP_BER,A:WLAN,A:Ad_hoc_WN}, the gain achieved by the proposed schemes with QPSK is about $5$ dB compared to the ideal RF front-end with BPSK\footnote{Notice that, for both schemes, the transmission rate is the same.}.\:For the IQI RF front-end with BPSK, an error floor is observed at an SER of $2.5\times 10^{-3}$, and the degradation caused by IQI on the average SER increases as the average SNR per bit gets larger. In the high $\rm{SNR}$ regime, a SER lower bound is observed. For example, for $16$-QAM, the SER floor is $5\times 10^{-2}$, which may not be acceptable in practice. Similarly, at a SER of $10^{-4}$, the SNR gain of IQSC with $16$-QAM is more than $15$ dB compared to the ideal RF front-end with QPSK.
Clearly, due to the MFD gain in \eqref{Eq:gamma_P}, IQSC achieves a better performance when compared to the ideal RF front-end. For the same $\rm{SNR}$ and transmission rate, systems employing IQSC can be far more reliable than single-input single-output (SISO) systems with ideal RF front-end, while they need less transmission energy to meet the same (practical) SER target.
In other words, IQSC systems can overcome the rate reduction by utilizing higher order modulation schemes.

\begin{figure}
\centering\includegraphics[width=0.95\linewidth,trim=0 0 0 0,clip=false]{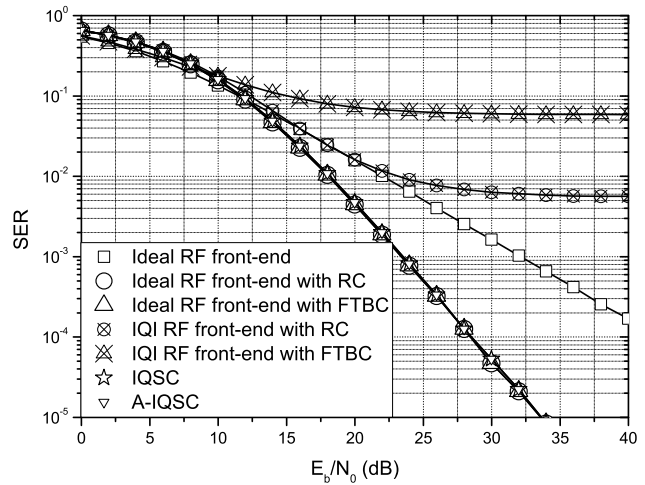}
\caption{Comparison of IQSC with RC and FTBC for $\mathrm{IRR}=20$ dB, $\phi=5^{o}$, and $16$-QAM.}
\label{fig:comparisson_RC}
\end{figure}

In Fig.~\ref{fig:comparisson_RC}, we illustrate the effect of IQI on the average SER of $16-$QAM, when RC and FTBC \cite{A:FTBC} are used. We compare IQSC and A-IQSC with RC and FTBC for the cases of an ideal RF front-end and an IQI RF front-end. Both RC and FTBC have the same rate as IQSC and A-IQSC. The IQI RF front-end with RC, as expected from \eqref{Eq:SNR_DD_upper_limit_2}, suffers from an error floor in the high-$\rm{SNR}$ regime. For the considered case, the SER floor is $5\times 10^{-3}$, which might not be acceptable in practice. Furthermore,the IQI RF front-end with FTBC, also exhibits an error floor in the high-$\rm{SNR}$ regime. For example, for $16-$QAM, the SER floor is $5.9\times 10^{-2}$, which again might not be acceptable in practice. On the other hand, due to the achieved MFD, IQSC and A-IQSC achieve a better performance than the IQI RF front-end with RC and FTBC, while they have a similar performance as the ideal RF front-end with RC and FTBC.

\section{Discussion and Conclusion}\label{S:Discusion&Conclusion}

In this section, we summarize the advantages and disadvantages of the two proposed IQSC schemes and discuss the new concept of MFD, highlighting our main findings.

\subsection{Merits and Drawbacks of IQSC}\label{S:Talk}

The proposed IQSC schemes are low-complexity countermeasures against any level of IQI, enabling their use in practical low-cost communication devices with unavoidable RF imperfections. IQSC and A-IQSC require neither extra transmit power nor any feedback from the RX, and their computational complexity is similar to Alamouti's space-time block coding scheme \cite{A:Alamouti}. Moreover, the proposed IQSC schemes (original and alternative) achieve second-order diversity without the use of extra antennas, which renders them far more reliable and power efficient than conventional SISO systems (with or without IQI compensation). Therefore, IQSC may find application in wireless systems, where low-cost, energy efficiency, low-complexity, and compactness of the communication devices are key design requirements. Notable examples are relaying systems and wireless sensors networks, where single-antenna nodes are physically limited in size, and have to operate without battery replacement for a long period of time~\cite{B:Cooperative_Communications,C:CooP_MIMO_WSN}.

In summary, both IQSC schemes have the following merits:
\begin{itemize}
\item They coordinate IQI in order to achieve two-fold diversity with a single antenna at TX and RX;
\item They are energy and/or bandwidth efficient compared to conventional SISO;
\item They do not require any type of feedback from the RX;
\item They are low-complexity schemes (their complexity is similar to that of Alamouti's space block coding scheme);
\item They provide interoperability and compatibility with existing wireless standards;
\item They support high-order modulation schemes avoiding an error floor in the SER due to IQI;
\item Their outage and SER performances are independent of the severity of the IQI.
\end{itemize}
The above reasons\,---and especially the last one---\,may allow the relaxation of hardware requirements and design specifications for future low-cost DCA communication devices.

The main drawback of the IQSC schemes is that they reduce the transmission rate by $50\%$. However, our results clearly demonstrate that uncompensated IQI severely degrades the error rate of conventional wireless communication systems especially for high constellation orders. Therefore, for high-order modulation schemes, performance may become unacceptable. On the other hand, when (A-)IQSC is employed, there is no such constraint. Hence, systems using (A-)IQSC can overcome the half-rate limitation by employing higher-order modulation schemes. As mentioned above, an obvious application of (A-) IQSC is the provision of diversity for low-cost remote units, which unavoidably suffer from RF imperfections, and have to operate without battery replacement for a long time.

\subsection{Conclusions}\label{S:End}

Two novel low-complexity techniques, namely IQSC and A-IQSC, were presented for significantly increasing the performance of single-antenna multi-carrier communication systems suffering from IQI. It was shown that the proposed schemes cannot only compensate the effects of IQI but they can even be beneficial for the system's performance due to the achieved MFD. In particular, IQSC (both original and alternative) outperforms the systems with uncompensated IQI at the RX, and the ideal RF front-end without IQI, when the same modulation order is employed. Moreover, it shows better performance compared to equal-rate repetition coding with uncompensated IQI, and the same performance as equal-rate RC with ideal RF front-end. Our results indicate that IQSC can be a promising technique for use in a variety of applications involving DCA low-cost devices.

\balance
\bibliographystyle{IEEEtran}
%\bibliography{IEEEabrv,References_v26}

\vspace{-0.0em}\begin{IEEEbiography}[{\includegraphics[height=1.25in,keepaspectratio]{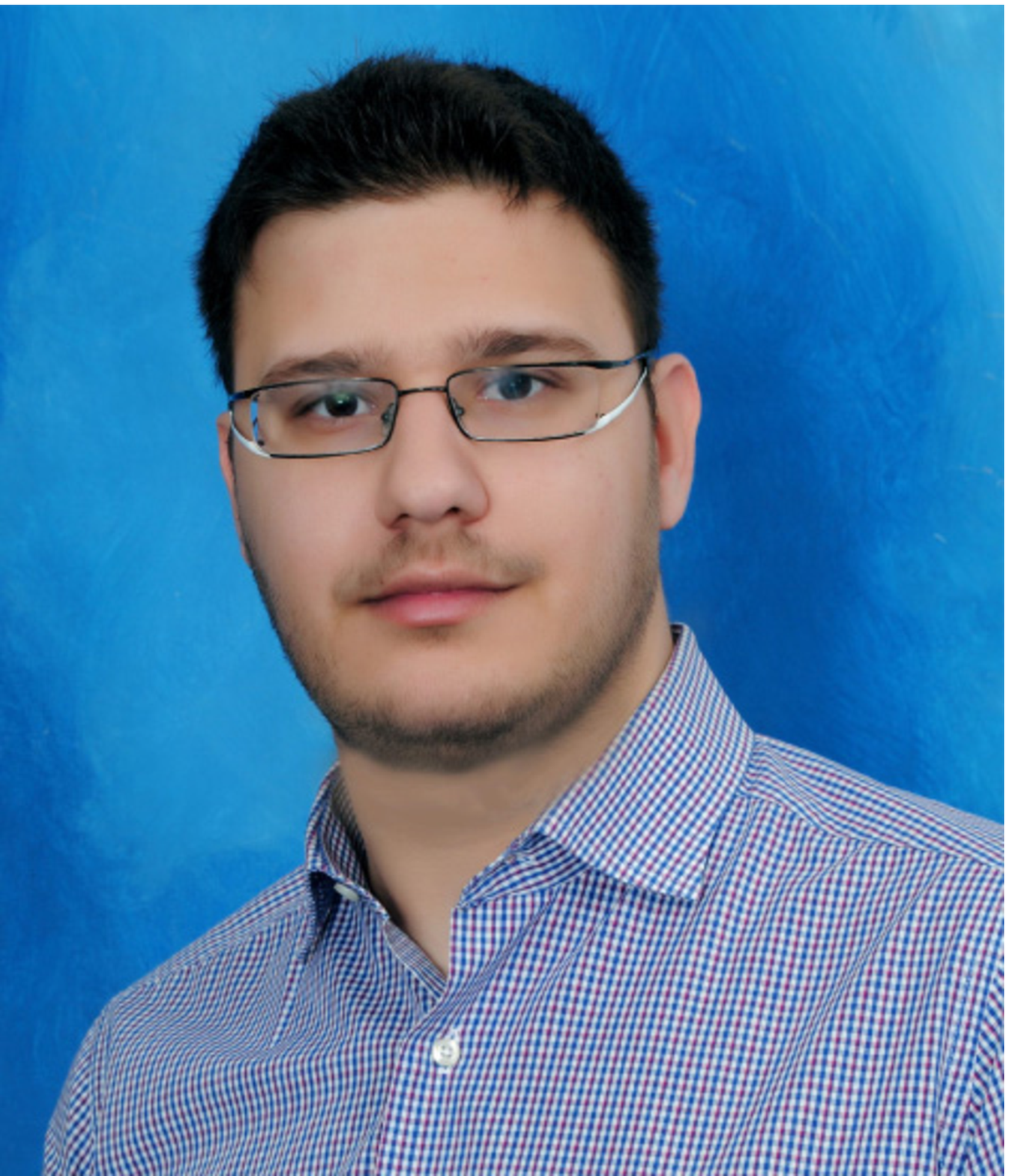}}]
{Alexandros--Apostolos~A.~Boulogeorgos} (S'11) was born in Trikala, Greece. He obtained the Diploma Degree (five years) in electrical and computer engineering from the Aristotle University of Thessaloniki, Greece, in 2012.
Since 2012, he has been working towards his Ph.D. degree at the Department of Electrical and Computer Engineering, Aristotle University of Thessaloniki. His current research interests are in the areas of fading channel characterization, cooperative communications, cognitive radio, interference management and
hardware-constrained~communications.
\end{IEEEbiography}\vspace{-0.0em}

\begin{IEEEbiography}[{\includegraphics[height=1.25in,clip,keepaspectratio]{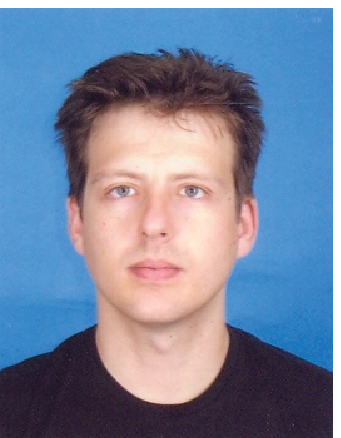}}]
{Vasileios M. Kapinas} (S'07--M'09) was born in Thessaloniki, Hellas (Greece). He holds a five-year diploma degree and a PhD both in electrical and computer engineering from the Aristotle University of Thessaloniki (AUTH). From 2001 to 2005 he was with AUTH as a Graduate Research Assistant at the Signal Processing and Biomedical Technology Unit. During the periods 2006-2009 and 2011-2014 he worked as a Research and Teaching Assistant at the Telecommunications Systems and Networks Lab of AUTH. From 2009 to 2011 he was with the TT\&C Systems and Techniques Section of the RF Payload Systems Division, European Space Research and Technology Centre (ESTEC), European Space Agency (ESA), The Netherlands.

Dr. Kapinas is currently a Postdoctoral Researcher with the Wireless Communications Systems Group, AUTH, focusing on next generation mobile and satellite communication systems. He is also a full member of the AIAA.
\end{IEEEbiography}\vspace{-0.0em}

\begin{IEEEbiography}[{\includegraphics[height=1.25in,clip,keepaspectratio]{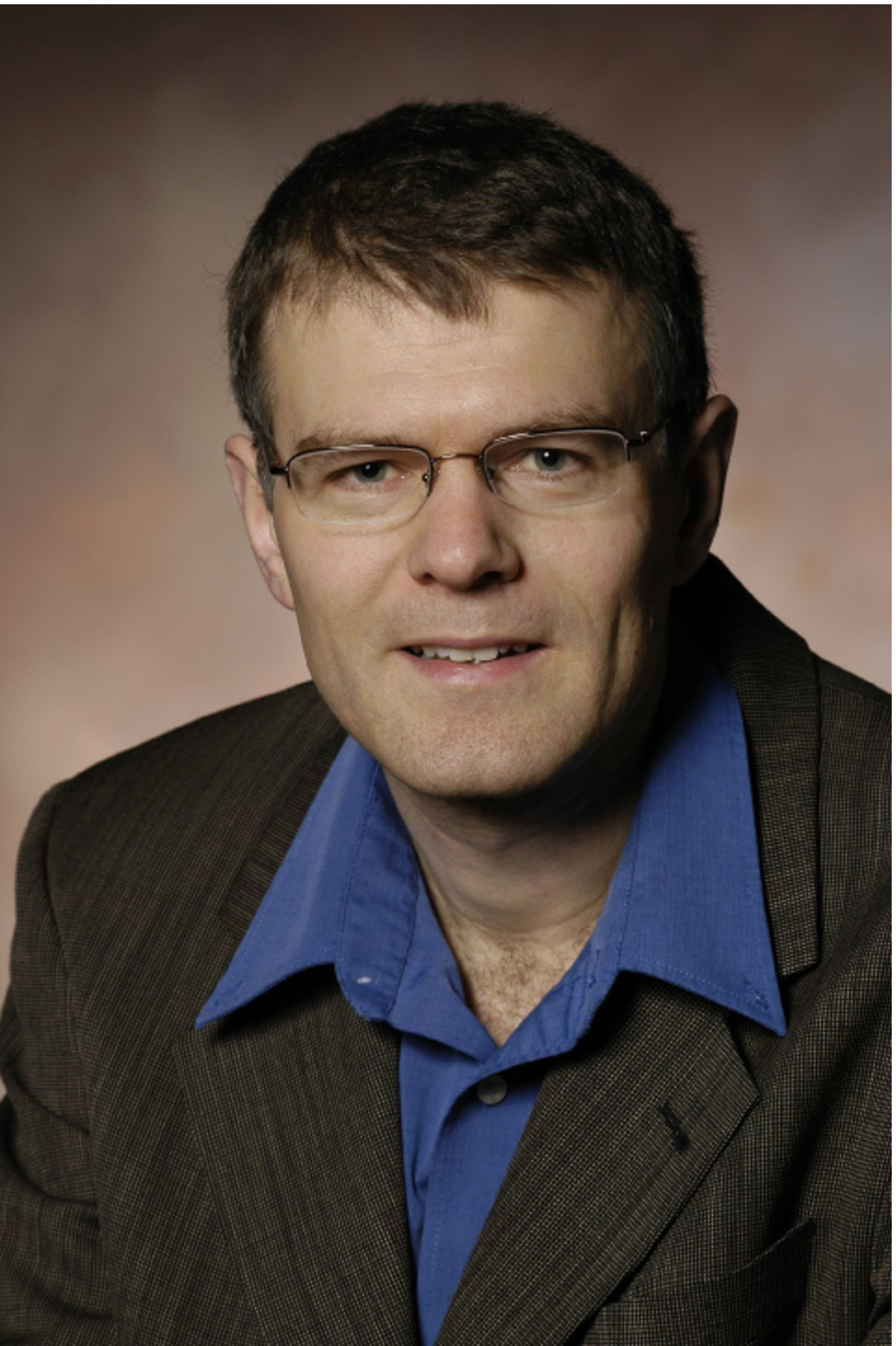}}]
{Robert Schober} (S'98, M'01, SM'08, F'10) was born in Neuendettelsau,
Germany, in 1971. He received the Diplom (Univ.) and the Ph.D. degrees in
electrical engineering from the University of Erlangen-Nuermberg in 1997
and 2000, respectively. From May 2001 to April 2002 he was a Postdoctoral
Fellow at the University of Toronto, Canada, sponsored by the German
Academic Exchange Service (DAAD). Since May 2002 he has been with the
University of British Columbia (UBC), Vancouver, Canada, where he is now a
Full Professor. Since January 2012 he is an Alexander von Humboldt
Professor and the Chair for Digital Communication at the Friedrich
Alexander University (FAU), Erlangen, Germany. His research interests fall
into the broad areas of Communication Theory, Wireless Communications, and
Statistical Signal Processing.

Dr. Schober received several awards for his work including the 2002 Heinz
Maier­Leibnitz Award of the German Science Foundation (DFG), the 2004
Innovations Award of the Vodafone Foundation for Research in Mobile
Communications, the 2006 UBC Killam Research Prize, the 2007 Wilhelm
Friedrich Bessel Research Award of the Alexander von Humboldt Foundation,
the 2008 Charles McDowell Award for Excellence in Research from UBC, a
2011 Alexander von Humboldt Professorship, and a 2012 NSERC E.W.R. Steacie
Fellowship. In addition, he received best paper awards from the German
Information Technology Society (ITG), the European Association for Signal,
Speech and Image Processing (EURASIP), IEEE WCNC 2012, IEEE Globecom 2011,
IEEE ICUWB 2006, the International Zurich Seminar on Broadband
Communications, and European Wireless 2000. Dr. Schober is a Fellow of the
Canadian Academy of Engineering and a Fellow of the Engineering Institute
of Canada. From 2012 to 2015, he was the Editor-in-Chief of the IEEE
Transactions on Communications. He currently serves as a Member-at-Large
on the Board of Governors of the IEEE Communication Society and as Chair
of the IEEE Transactions on Molecular, Biological and Multiscale
Communications.
\end{IEEEbiography}

\begin{IEEEbiography}[{\includegraphics[height=1.25in,clip,keepaspectratio]{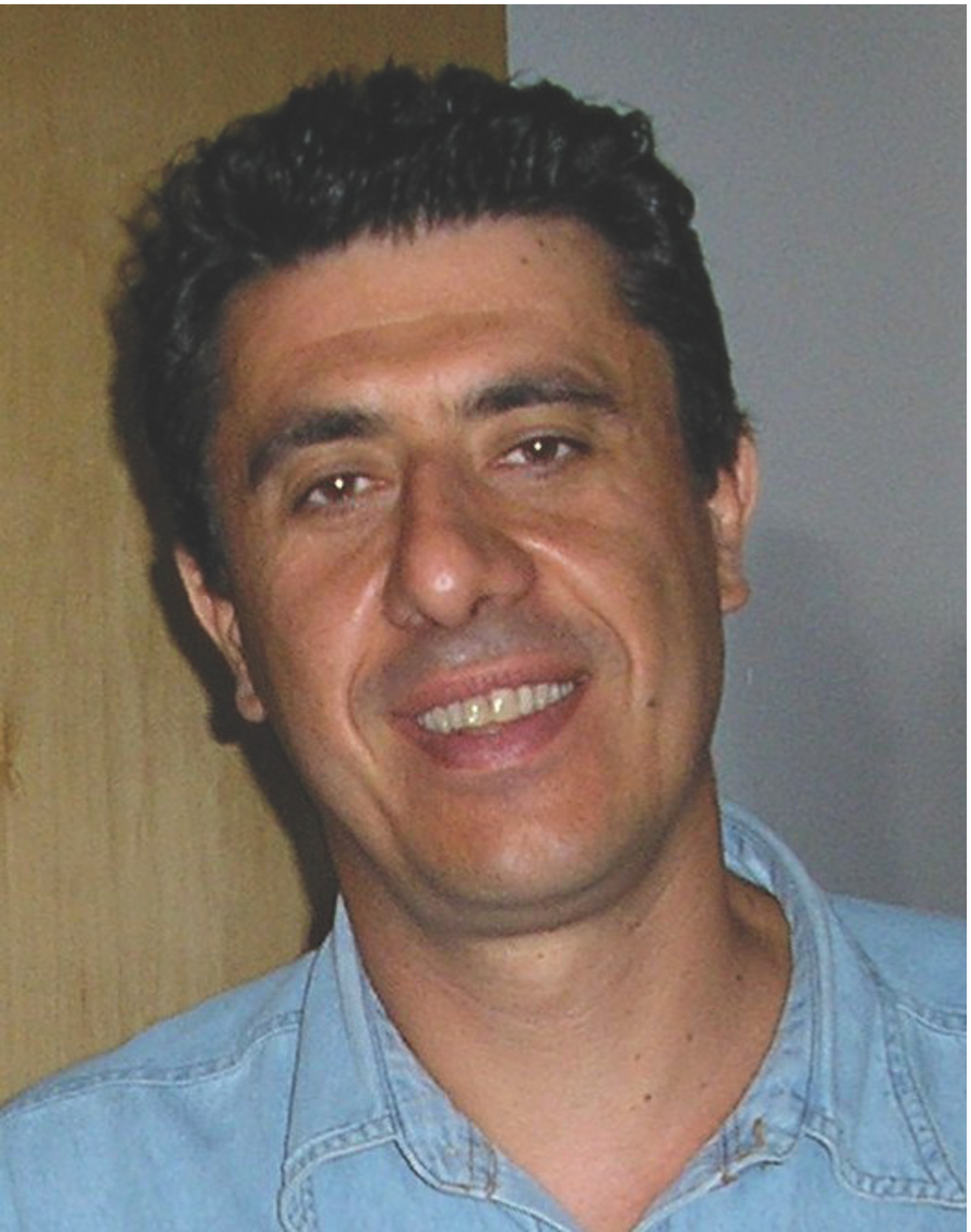}}]
{George K. Karagiannidis} (M'96-SM'03-F'14) was born in Pithagorion, Samos Island, Greece. He received the University Diploma (5 years) and PhD degree, both in electrical and computer engineering from the University of Patras, in 1987 and 1999, respectively. From 2000 to 2004, he was a Senior Researcher at the Institute for Space Applications and Remote Sensing, National Observatory of Athens, Greece. In June 2004, he joined the faculty of Aristotle University of Thessaloniki, Greece where he is currently Professor in the Electrical \& Computer Engineering Dept. and Director of Digital Telecommunications Systems and Networks Laboratory.
His research interests are in the broad area of Digital Communications Systems with emphasis on Wireless Communications, Optical Wireless Communications, Wireless Power Transfer and Applications, Molecular Communications, Communications and Robotics and Wireless Security.

He is the author or co-author of more than 400 technical papers published in scientific journals and presented at international conferences. He is also author of the Greek edition of a book on ``Telecommunications Systems'' and co-author of the book ``Advanced Optical Wireless Communications Systems'', Cambridge Publications, 2012.

Dr. Karagiannidis has been involved as General Chair, Technical Program  Chair and member of Technical Program Committees in several IEEE and non-IEEE conferences. In the past he was Editor in IEEE Transactions on Communications, Senior Editor of IEEE Communications Letters, Editor of the EURASIP Journal of Wireless Communications \& Networks and several times Guest Editor in IEEE Selected Areas in Communications. From 2012 to 2015 he was the Editor-in Chief of IEEE Communications Letters. He is a Honorary Professor at South West Jiaotong University, Chengdu, China.

Dr. Karagiannidis has been selected as a 2015 Thomson Reuters Highly Cited Researcher.
\end{IEEEbiography}
\end{document}